\documentstyle[12pt,infnprep,epsfig]{article}
\newcommand{\Header}{
  \begin{tabular}{rl}
  \hspace{-2.5cm}\includegraphics{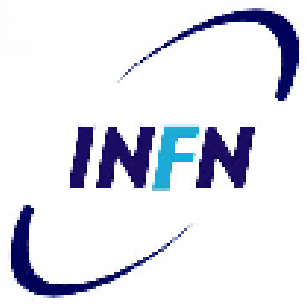} &
    \renewcommand{\arraystretch}{0.5}
  \begin{tabular}{r}
      {\hspace{1cm}~\LARGE\sffamily ISTITUTO NAZIONALE DI FISICA NUCLEARE}\\
      \\
    \end{tabular}
  \end{tabular}
\begin{center}
    {\Large\sffamily Sezione di Milano Bicocca}\\
\end{center}
    \renewcommand{\arraystretch}{1}
\rule{16.0cm}{0.09mm}
\vskip 1.0cm
  \begin{flushright}
      {\underline{\bf INFN/AE-10/1 }}\\    
      {\small\bf 22 Gennaio 2010} \\      
  \end{flushright}
}
\begin{document}
\begin{titlepage}
\title
{ \Header  \large \bf The design and commissioning of the MICE upstream time-of-flight system.}
\vskip 1cm
\author{
R.~Bertoni$^1$, A.~Blondel$^2$, M.~Bonesini$^{1,*}$,
G.~Cecchet$^3$, A.~de Bari$^3$, J.S.~Graulich$^2$,\\
Y.~Kharadzov$^4$, M.~Rayner$^{1,+}$, I.~Rusinov$^4$, 
R.~Tsenov$^4$, S.~Terzo$^1$, V.~Verguilov$^4$ \\
{\it ${}^{1}$ INFN -- Sezione di Milano Bicocca,  
Milano, Italy} \\
{\it ${}^{2}$ Section de Physique, Universite` de  Geneve,  
Geneve, Suisse} \\
{\it ${}^{3}$ INFN -- Sezione di Pavia, Dip. di Fisica Nucleare 
e Teorica,  Pavia, Italy} \\
{\it ${}^{4}$ Department of Atomic Physics, St. Kliment Ohridski University, 
 Sofia, Bulgaria}}
\maketitle
\baselineskip=14pt

\vskip 0.5cm
\begin{abstract}
In the MICE experiment at RAL the upstream time-of-flight detectors are used
for particle identification in the incoming muon  beam, for the experiment
trigger
and for a precise timing ($\sigma_t \sim 50$ ps) with respect to the
accelerating RF cavities working at 201 MHz.
The construction of the upstream section of the MICE
time-of-flight system and the tests done to characterize
its individual components are shown.
Detector timing resolutions $\sim 50-60$ ps were achieved.
Test beam performance and preliminary results obtained with beam at RAL
are reported.
\end{abstract}

\center{\it (submitted to Nuclear Instruments and Methods A) }

\vspace*{\stretch{2}}
\begin{flushleft}
  \vskip 0.6cm
{${}^{*}$  Corresponding~author: M.~Bonesini, 
E-mail~address:~maurizio.bonesini@mib.infn.it} \\
{${}^{+}$ permanent address:  Department of Physics, Oxford University, UK},
\end{flushleft}
\begin{flushright}
  \vskip 1cm
\small\it Published by {\bf SIS-Pubblicazioni}\\
Laboratori Nazionali di Frascati
\end{flushright}
\end{titlepage}
\pagestyle{plain}
\setcounter{page}2
\baselineskip=17pt

\begin{figure}[htb]
\begin{center}
\includegraphics[width=0.9\textwidth]{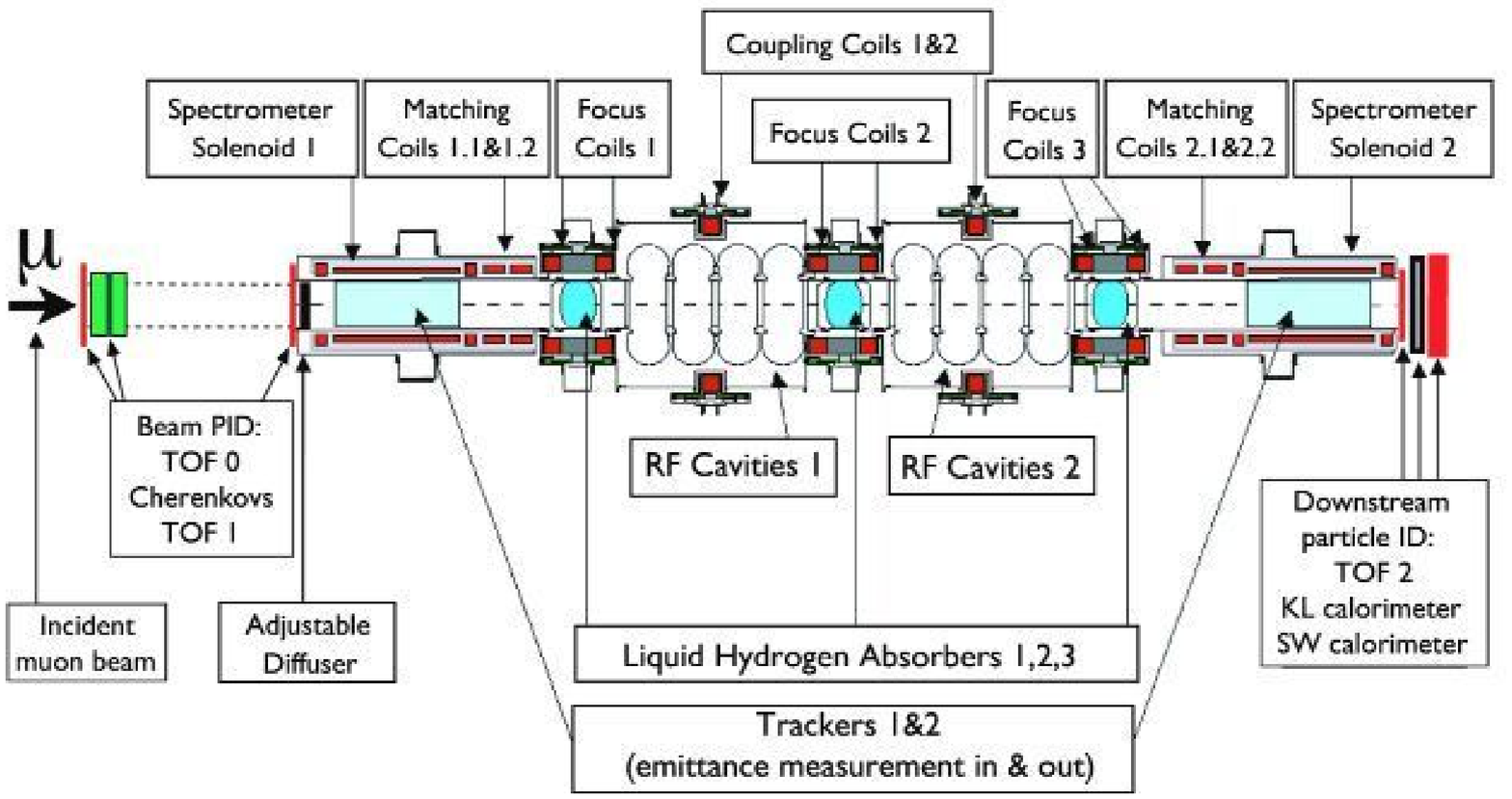}
\caption{ 
2-D layout of the MICE experiment at RAL (not in scale).
The muon beam from ISIS (140-240 MeV/c central momentum,
tunable between 1-10 $\pi \cdot $ mm rad input emittance)
enters from the left. The cooling section
is put between two magnetic spectrometers and two TOF stations
(TOF1 and TOF2) to measure particle parameters. The input beam composition is
determined by two Aerogel Cherenkov counters and the two upstream TOF 
detectors (TOF0 and TOF1).}
\label{fig-mice}
\end{center}
\end{figure}

The MICE experiment \cite{mice} at RAL 
(see figure \ref{fig-mice} for a schematic layout)
aims at a
systematic study of a section of a cooling channel 
of a neutrino factory ($\nu F$) \cite{kosharev}.
The 5.5 m long cooling section consists of three liquid Hydrogen 
absorbers and eight 201 MHz RF
cavities encircled by lattice solenoids.

Different neutrino factory  designs require a muon cooling factor from 2 to 16,
over a $\sim 100$ m distance. 
For a cooling section prototype of affordable size,
a cooling factor $\sim 10 \%$ at most may be expected.
A precision of $\sim 10 \%$ on the design of the 
whole cooling channel implies 
emittance measurements at a level of $0.1 \%$ on
the cooling cell prototype,
thus excluding conventional emittance measurement methods,  
that have errors around $10 \%$.

A method based on single particle measurements
has been envisaged, to obtain such a level of precision.
Particles are measured before and after the cooling section
by two magnetic spectrometers complemented by time-of-flight (TOF)
 detectors.
For each particle x, y, t, 
$p_x$, $p_y$, $E$ coordinates
are measured.
In this way, for an ensemble of N particles, the
input and output emittances may be  determined  accurately.

\section{The upstream MICE time-of-flight system}
In the MICE experiment, precision timing measurements are required to 
relate the time of the incoming beam muons to the phase of the 
accelerating field in each RF 
cavity and simultaneously for particle identification (PID) by 
a TOF method. 
Three time-of-flight detectors (TOF0, TOF1, TOF2) are foreseen. The last two (TOF1 and TOF2) are 
at the entrance and the exit of the MICE cooling channel; the first one (TOF0) instead is 
placed about 10 m  upstream of its entrance. 
Figure \ref{fig-mice} shows a layout of the full MICE
cooling channel with the foreseen positions of the TOF detectors.
The upstream TOF detectors (TOF0, TOF1) must separate the pion 
contamination of the muon beam at low momenta (below $\sim 210$ MeV/c)
and are used for the 
experiment trigger. All TOF detectors are used to determine
the time coordinate ($t$) in the measurement  of the emittance. 

The TOF stations share a common design based on two planes of 
fast one-inch scintillator 
counters along X/Y directions (to increase measurement redundancy) read at both edges by 
R4998 Hamamatsu fast photomultipliers (PMTs)~\footnote{one-inch linear focused PMTs, typical gain $G \sim 5.7 \times 10^{6}$ 
at -2250 V, risetime 0.7 ns, transit time spread (TTS) $\sim 160$ ps}. In the upstream section, 
the TOF0 planes cover a
$40 \times 40 \ cm^2$ active area, while the TOF1 planes cover  a 
$42 \times 42 \ cm^{2}$  active area.
The counter width is 4 cm in TOF0 and 6 cm in TOF1. Time calibration of individual
counters has been done with impinging beam particles by using the detector 
X/Y redundancy.
In addition a fast laser calibration system is foreseen for monitoring.

To determine the timing with respect to the RF phase to a precision of $5^0$ 
a detector resolution
$\sim 50$ ps is needed for TOF0, while to  allow a $99 \%$ 
rejection of pions in the incoming muon beam, a resolution better than
 $\sim 100$ ps is sufficient for the TOF measurement 
between TOF0 and TOF1. 
The resolution in the TOF measurement between detectors $i$ and $j$ 
is expressed as:

\begin{equation}
 \sigma_{TOF_{i,j}} = \sqrt{ \sigma_{T_i}^2 + \sigma_{T_j}^2 + 
\sigma_{calibr}^2} 
\label{tof:sigma}
\end{equation}

where $\sigma_{T_i} (\sigma_{T_j})$ is the i-th (j-th) TOF station   time resolution 
and $\sigma_{calibr}$ is the resolution of the calibration system. Having two independent 
measurements from each TOF stations (due to the X/Y redundancy) $\sigma_{T_{i}}$ is given 
by $\sigma_t/\sqrt{2}$, where $\sigma_t$ is the intrinsic counter timing 
resolution. 

Taking into account also the calibration errors, this imply a conservative 
requirement for single detector timing resolution $\sigma_{T} \sim 50-60$ ps 
and a resolution of
the calibration procedure $\sigma_{calibr} \sim$ 50 ps.

\subsection{Working conditions of the time-of-flight detectors inside MICE}
In the MICE experiment the TOF detectors have to work with 
high incoming particle rates (up to 1.5 MHz), 
high magnetic fringe fields 
 from the tracking solenoids with {$ \bf \mid B \mid$} up to $\sim 1300$ G
(only for TOF1 and TOF2)
and a high level of RF noise from 
the cooling channel.

From beamline simulations  and the expected 
beam widths at the TOF0 and TOF1 detectors positions ($\sigma_{x,y} \sim 3.3-4$ cm) 
rates up to about 0.5 MHz must be sustained 
by single PMTs.

\begin{figure}[hbt]
\begin{center}
\includegraphics[width=5.2cm]{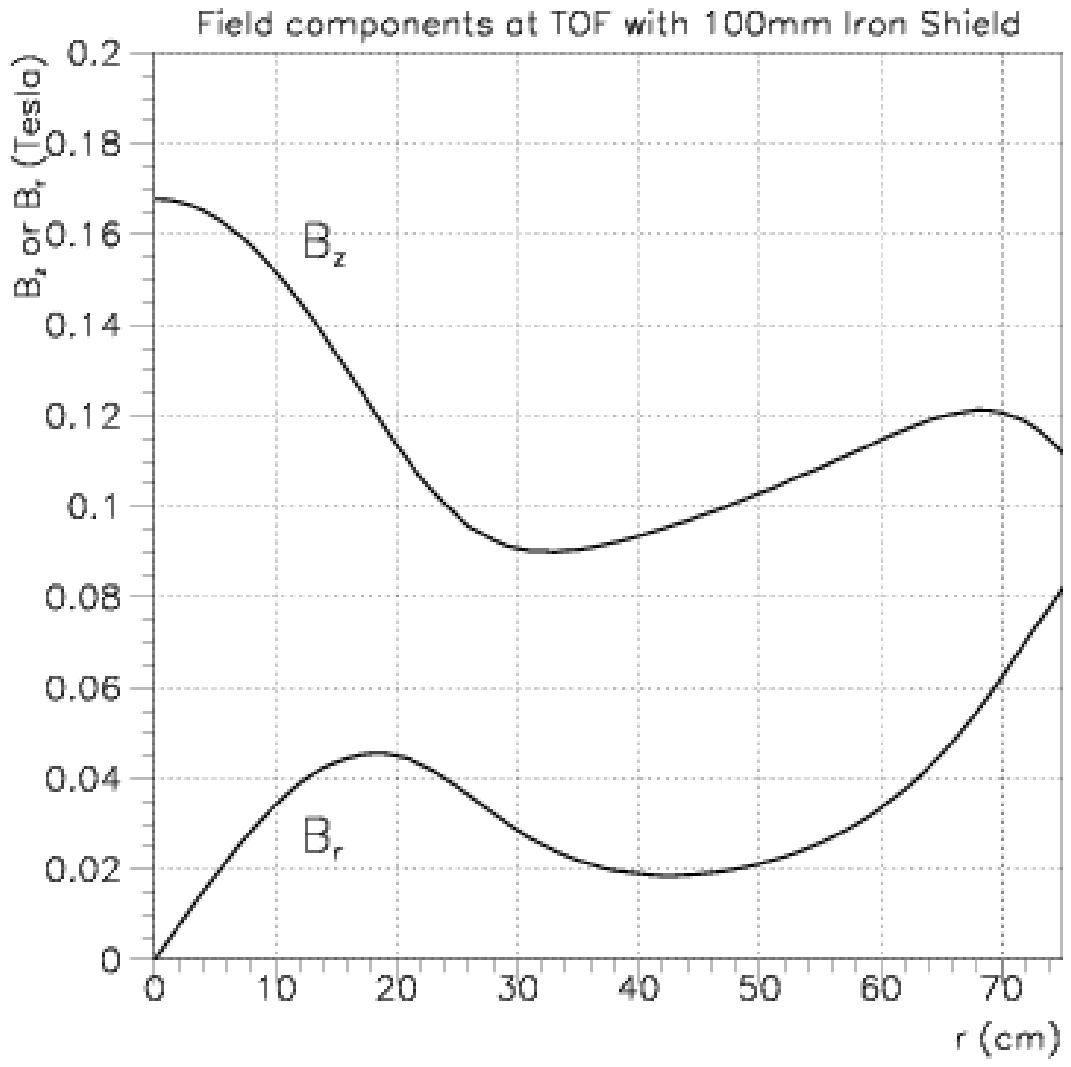}
\includegraphics[width=6cm]{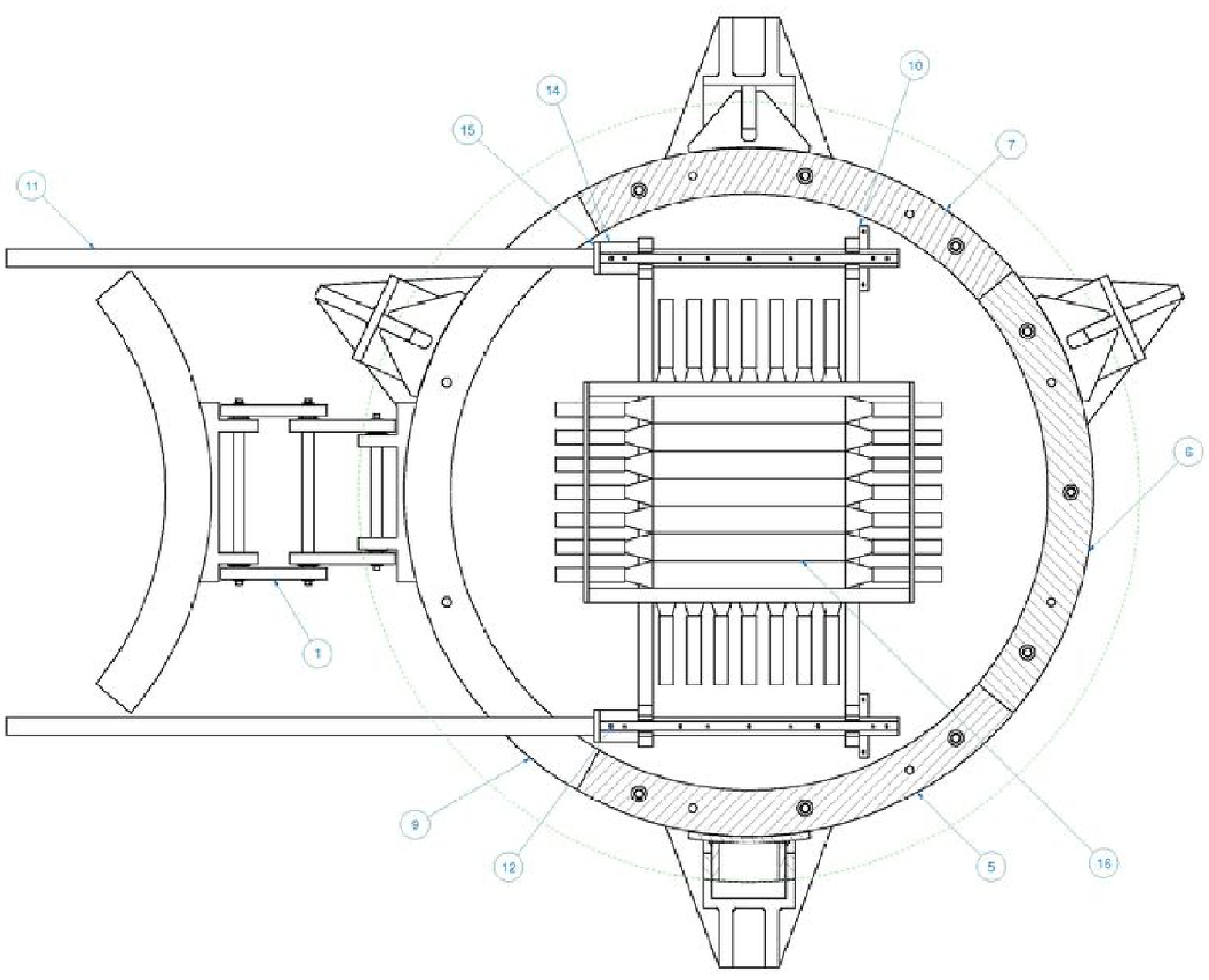} 
\caption{Left panel: longitudinal $B_{\|}$ and orthogonal $B_{\bot}$ 
components of the residual magnetic field, as a function
of the radial distance $r$ from the beam axis at the position of TOF1 or TOF2,
after a 100 mm annular shielding plate \cite{cobb}. Right panel: 
magnetic shielding cage for TOF1 (front view). 
The TOF1 detector is shown inside the 
shielding cage with sliding rails at top/bottom to extract the detector
and  the extraction brackets at the left side \cite{gregoire}.}
\label{fig:tof1}
\end{center}
\end{figure}

Due to the low residual magnetic  field  produced by
  the last  quadrupole of the beam channel 
 in the vicinity of the TOF0 detector ($\leq 50$ Gauss), conventional PMTs 
with an elongated mu-metal shielding 
(extending 30 mm beyond the photocathode surface) may be used
(see later for details). The other 
two stations (TOF1 and TOF2) 
will work instead inside the high residual magnetic field 
of the spectrometer  solenoids, 
that is only partially 
shielded by a 100 mm iron annular plate. The left panel of 
figure \ref{fig:tof1} shows the 
residual longitudinal $B_{\|}$ and orthogonal $B_{\bot}$ 
components of the magnetic field at the position of TOF1 and TOF2 detectors, as 
computed with a 2D Tosca \cite{tosca} or COMSOL \cite{comsol} 
calculation~\footnote{ 
3D Tosca calculations were redone and results were found compatible
\cite{gregoire1}}. 
Because orthogonal components (up to $\sim 1200$ Gauss) and longitudinal components (up to $\sim 400$ Gauss) 
of the fringe magnetic fields must be shielded, 
a local or a global magnetic shielding for
TOF1 and TOF2 detectors has to  be envisaged. For conventional 
PMTs~
\footnote{ Other solutions, based on PMTs for high magnetic 
fields such as Hamamatsu R5505-70, have been studied, but later they have
been abandoned for their much higher cost \cite{bonesini2}}
 the most difficult component to be shielded is the one along the PMT's axis. 
Orthogonal components can be more easily shielded.
A global cage  bolted to the 
annular return plate of the nearby
spectrometer solenoid will be used for TOF1. 
This is shown in the right panel of figure \ref{fig:tof1} with the 
relevant mechanical details.
 
As computed with a 3D Tosca calculation \cite{gregoire1} the residual
field inside the shielding cage is below a few Gauss: a value well tolerable
by the  R4998 PMTs with a 1 mm $\mu-$metal shielding. 
The solution, albeit elegant, has the drawback of the need of 
a quite complicate 
extraction mechanism to allow
access to the detector inside the inner volume of the shielding cage.

\section{Detector construction}
The structure of TOF1, inside the shielding cage, is shown in the right
panel of figure \ref{fig:tof1}. TOF0 has a similar crossed X/Y structure. 
Each scintillator slab, after a straight Polymethyl methacrylate (PMMA)
 lightguide, is read at the
two edges by a fast R4998 PMT. 
Scintillator counters have been assembled in-house starting from 
DTF (diamond tool finished) scintillator bars from Bicron, to which PMMA 
light-guides  have been glued with BC-600
 optical cement.
A simple design with flat fish-tail PMMA lightguides, instead of 
tilted ones
(to reduce the influence of magnetic field) or Winston cones, has
been chosen to optimize the timing detector resolution (favouring
the collection of straight light) and to allow an easy mechanical 
assembly. 
The chosen design of the lightguides has been checked with a dedicated
simulation program \cite{guideit}.
Wrapping and assembly has been realized with total tolerances less 
than  1 mm
for each individual counters of the TOF0/TOF1 planes. 
The final choice of wrapping is  aluminized mylar + 
black PVC covering. The light-tightness of the  covering material
has been tested measuring the transparency of a small sample inside a
spectrophotometer~\footnote{model JASCO V-530 UV/VIS}. 
The optical contact between the end of the lightguide collar  and the PMT 
photocathode is assured by silicone elastomers~\footnote{Bicron 
one-inch BC-634 optical pads}. 


For the scintillator material, different options have been considered
(see table 1 for more details). The Bicron BC-420 scintillator 
has been retained as a choice
for TOF0, while BC-404 have been used for TOF1 and TOF2. 
In spite of small additional problems for the choice of lightguide
material (high quality UVT plexiglas, instead of commercial UVA plexiglas, as
the scintillation emission peak is around 390 nm), BC-420 was  expected to give
slightly  better timing performances than BC-404 and was thus considered
the optimal choice for TOF0.
\begin{table*}[hbtp]
\begin{center}
{\footnotesize
\begin{tabular}{|c|c|c|c|c|c|c|}
\hline
              & BC-408   &  BC-404  & BC-420 & EJ-204 & EJ-230 & UPS-95F \\ \hline
$\lambda^{max}_{emission}$ (nm) & 425      &  408     & 391    & 408    & 391& 390        \\
$\lambda_{att}^{bulk}$ (cm) & 380 & 160 & 110 & - & $\sim 100$ & - \\
Light output $\%$ Anthr.    & 64& 68 & 64 & 68 & 64 & 39-45                        \\
decay const. (ns) & 2.1& 1.8 & 1.5 & 1.8 & 1.5 & 1.2              \\
risetime (ns) &  0.9 & 0.7 & 0.5 & 0.7 & 0.5 & 0.7           \\
pulse width (FWHM ns)      & & & & 2.2 & 1.3 & -        \\
\hline
\end{tabular}
\caption{Main properties of considered scintillator for TOF0/TOF1 counters,
from Bicron, Eijlen Technologies and Amcrys-H. 
BC-420 and EJ-230 (BC-404 and EJ-204) have similar composition.}
\label{tab1}}
\end{center}
\end{table*}

\begin{figure}[hbt]
 \begin{center}
 \includegraphics[width=12cm]{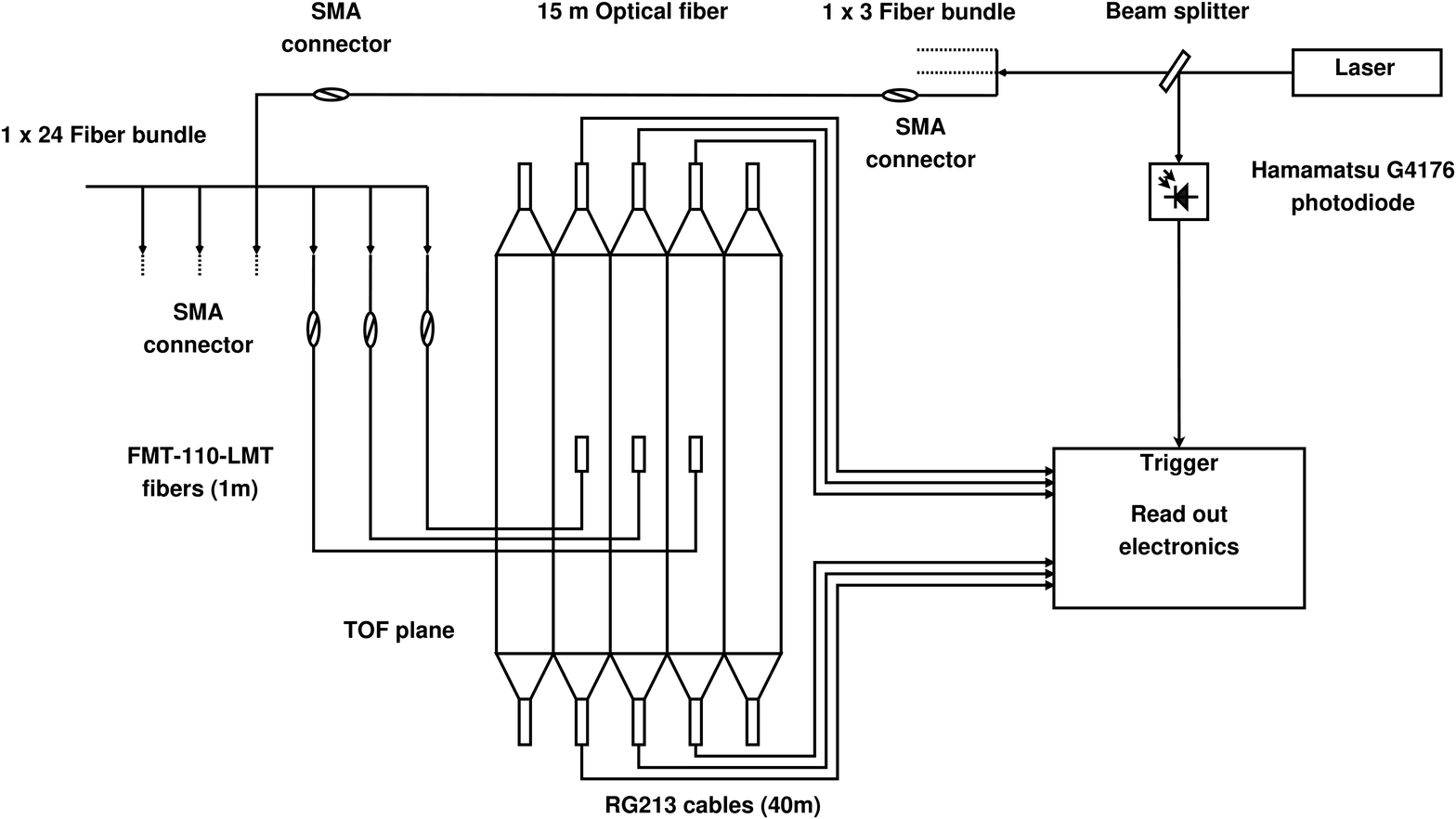}
 \caption{ Layout of the fast laser calibration system.} 
 \label{fig-laser}
 \end{center}
 \end{figure}

Time calibration of individual counters has been  done with 
impinging beam particles, using the X/Y redundancy of 
TOF detectors (see later for details). 
A fast laser calibration system, as in the HARP experiment
large TOF wall detector \cite{bonesini}
is foreseen for time calibration and monitoring 
(see figure \ref{fig-laser} for details). The laser light is beam split
to a fast Hamamatsu G4176 photodiode, giving
the system START, and is injected into a bundle 
of fibers that transmit the pulse to the different scintillator counters.

Studies are under way to provide an economic and stable fast laser source.
To reduce launch problems, IR monomode Corning SMF-28 fibers, 
that for blue or green light behave as a ``limited'' number of modes fiber,
will be used.

The fiber bundle will be realized with a $1 \times 3$ fused-silica splitter
followed by three $\sim 15$m long  fibers going each one to a 
$1 \times 24$ fused-silica splitter. The splitters, realized by OZ 
Optics~\footnote{OZ Optics Ltd., Ottawa, Canada} with Corning SMF-28 fibers,
have splitting ratios with relative differences less than 
$\pm 10 \%$ (rms) for the 20 (14) fibers
to be used for TOF0 (TOF1).

Laser light will be injected 
at the center of each counter by a total reflection prism, after a 1 m long
multimode (MM) 
fiber~\footnote{FT-110-LMT from 3M, with core diameter $110 \mu$m and
typical attenuation $20 dB/km$ at 500 nm} that convey the laser pulses. 
The total reflection prism and the fiber holder are glued inside a 
black PVC cap 
with black silicone~\footnote{Dow Corning 732 sealant} to ensure 
light-tightness.

\subsection{Electronics readout}

\begin{figure}[hbt]
\begin{center}
\includegraphics[width=\linewidth]{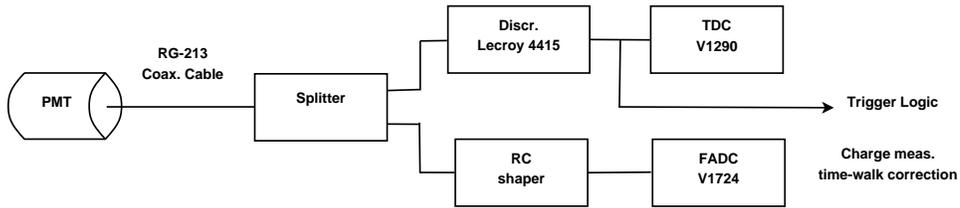}
\caption{Schematic layout of the MICE TOF front end electronics }
\label{front_end}
\end{center}
\end{figure}

A schematic layout of the front-end electronics is shown in  
figure \ref{front_end}. The PMT signal is split to a time-to-digital (TDC)
line and a sampling flash analog-to-digital converter (FADC)
line for time-walk-corrections. The PMT pulse measurement scheme 
is designed to meet the high input event rate requirements in MICE, that
demand electronic modules with conversion times better than $1 \mu$s and
a $\sim 1000$ events buffer. 

The negative signal from the PMT, after a 40 
m long RG-213 cable, passes through a 
passive ($50 \%-50 \%$) splitter and then  is sent to a RC shaper and to 
a leading edge discriminator LeCroy 4415. 
Fast timing cables RG-213 rather than conventional RG-58 cables have been
used to reduce signal distortion. As measured in \cite{harp-tof}, RG-213 cables
have a better stability as a function of temperature: $30 ppm/{}^{0}C$  single
channel temperature variation that is three times better than standard 
RG-58 cables. This reduces only to a few $ppm/{}^{0}C$ when considering 
the relative channel to channel variation. 
Before installation at RAL, the delays $\delta_{j}$ introduced by
the signal cable lengths have been individually measured. 

The RC shapers and splitters are specifically designed and produced for the requirements of the MICE experiment. 
A short acquisition time for a pulse is achieved by using 
a FADC~\footnote{ 
CAEN V1724 FADC with 100 MS/s maximum sampling rate}. The shaping circuit is used to effectively extend the duration 
of the short PMT pulse, so that it could be finely sampled in successive time points by the ADC. Software processing 
of the digitized pulse yields its amplitude or the area  values that are
 needed for charge measurement and time-walk correction. 
The passive splitter is designed to match the impedances of the 50 Ohm coaxial cable, coming from the PMTs, 
with the  120 Ohm  impedance of the Lecroy 4415 leading edge
discriminators and shaper inputs. The shaper circuit provides 4-stages of low-pass 
filtering and amplification of the input pulses. The time-constant of the filter stages is selected around 30 ns, 
which gives effective stretching of the 5 ns PMT pulses up to 400 ns. In this way the acquisition of a 
pulse can take place within the available $1 \mu$s time interval, 
providing several tens of sampled points along 
the pulse profile. The signal is DC-coupled throughout the whole shaper circuit. This provides baseline 
insensitivity to pulse rate variations. The DC gain can be selected from several predefined values by jumper settings. 
Sixteen shaper channels are organized in one NIM module, each channel having individual gain, offset voltage 
and polarity adjustments. 

Similarly, the splitter board is also a 16-channel unit. It is mounted directly on the front 
panel of the shaper module. A twisted-pair flat cable is used to connect 
the splitter with the discriminator module.

After the discriminator a fast CAEN V1290 TDC is used to provide timing measurements.
The V1290 is a multihit/multievent VME TDC that can detect hits rising/falling edges and work
in continuous storage mode with a $32K \times 32$ bits deep outer buffer.
A 25 ps least significant bit (LSB)  
couples to a 5 ns double hits resolution. 
The CAEN V1290 TDC has a differential non-linearity (DNL) 
of 2.8 LSB and an integrated non-linearity (INL) of 15 LSB, as reported
in \cite{tintori}. 

\begin{figure}[hbt]
\begin{center}
\includegraphics[width=6cm]{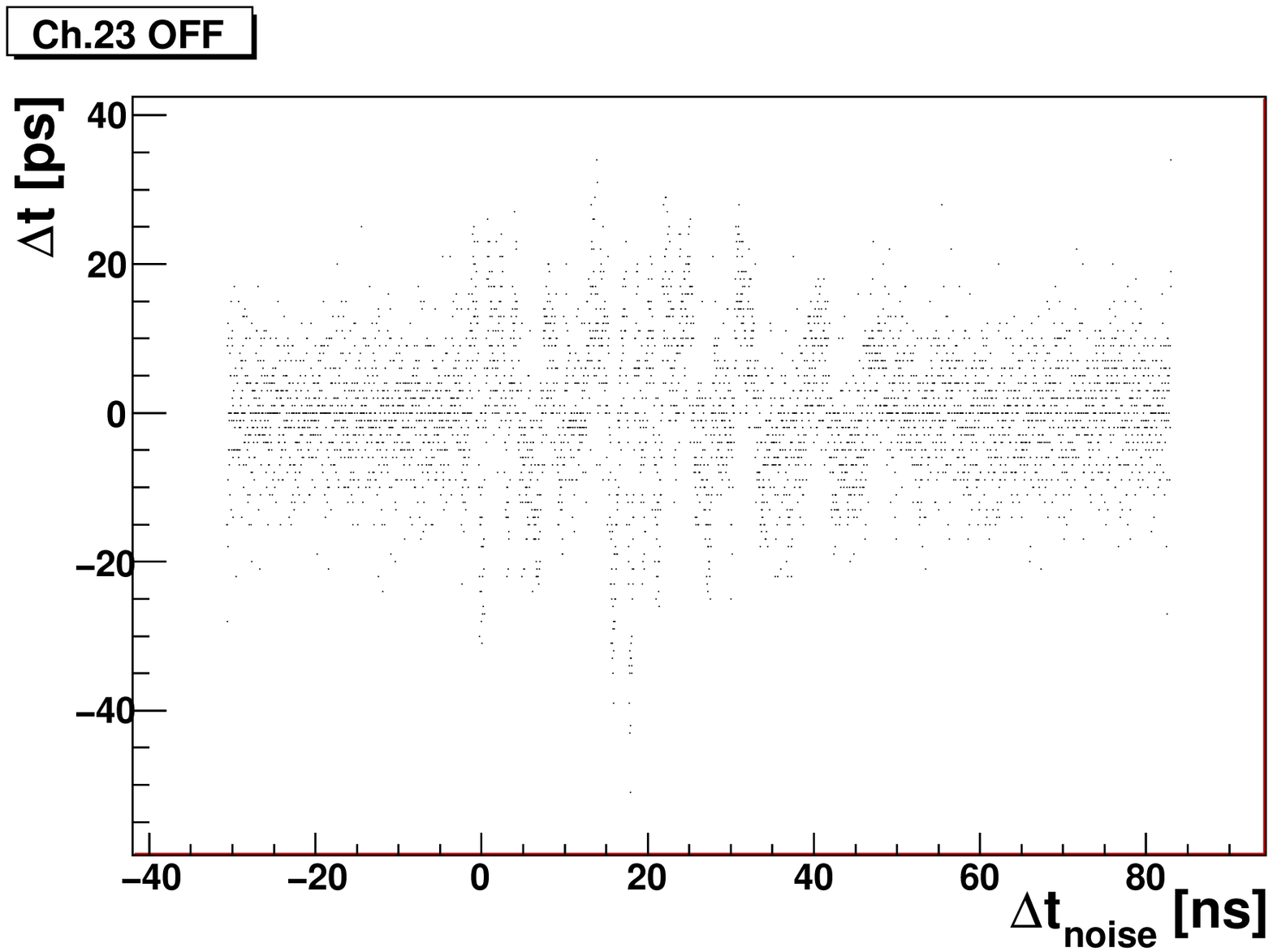} \\
\includegraphics[width=6cm]{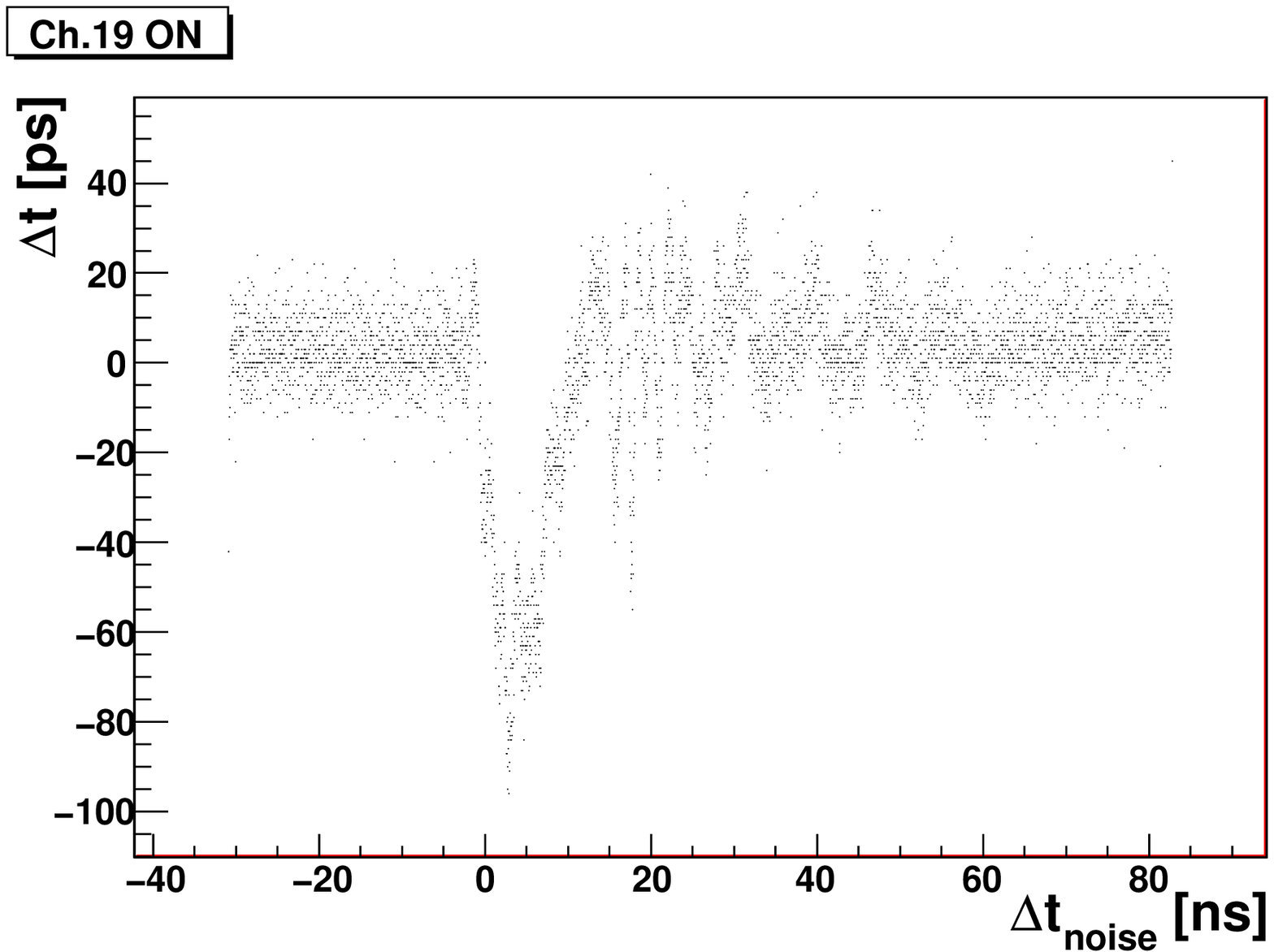}
\includegraphics[width=6cm]{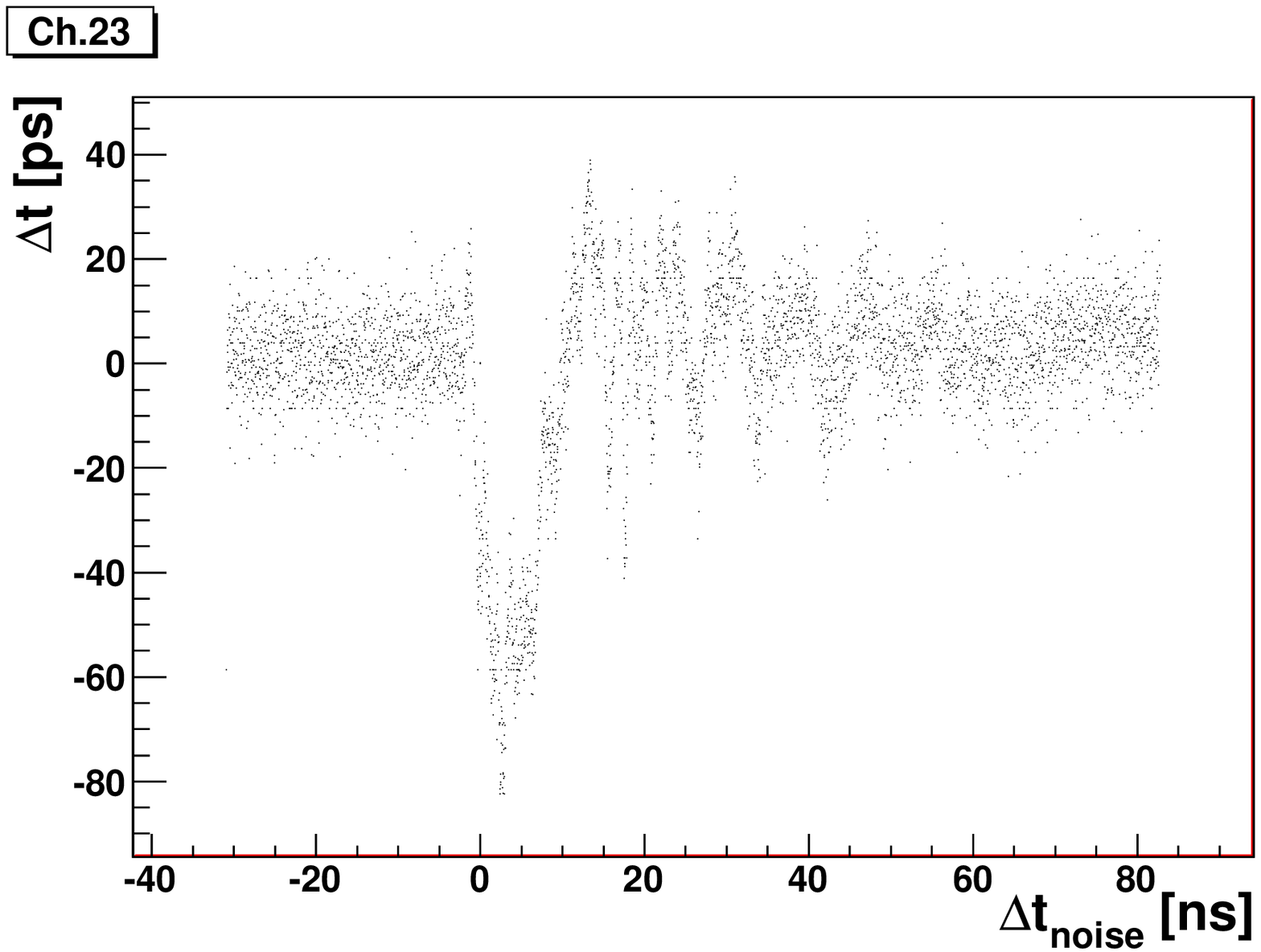}
\caption{Scatter plot of the time difference between the two split stop
signals versus the time difference between the disturbing noise and one stop
signal. The disturbing signal is coming to the same board of one of the
stop signal in the two bottom panels, while it is coming to another board 
in the top panel (no effect seen).}
\label{fig:noise}
\end{center}
\end{figure}

For timing measurements, a relevant  problem is 
given by the cross-talk due to channel-to-channel coupling in the same
TDC electronic board. The 32 channels of a V1290A TDC are grouped into
four separate electronic boards. 
A fixed start-stop measurement, with the stop signal split
into two different channels belonging to different electronic boards, 
was implemented. One of the stop signals
was then disturbed by a pulse with a sweeping delay with respect to it,  
coming to another channel of the same (different) board to which
the STOP line is connected. Figure \ref{fig:noise}
shows the difference of the two stop signals (peaking at zero in absence of
external noise) as a function of the delay of the external noise with respect
to  one of the stop signals. The top panel shows the case of the signal
coming to a channel of another board (where no cross-talk is expected), 
while the other two panels show 
the case when the noise is coming to a channel of the same board of the 
stop signal.
Coming to a different board the effect is less than 1 LSB, while coming
to a channel of the same board the effect reaches 3 LSB in a time window less
than 20 ns.  

With an incoming particle rate of less than 1 Mhz for single counting,
this poses no serious problems for the timing measurement.

\section{Tests of single detector components}
Single components of the TOF detectors were individually
characterized for optimal performances. In particular,
extensive studies were done on the fast Hamamatsu R4998 PMTs
and to choose the most suitable  scintillator material.

\subsection{Tests on Hamamatsu R4998 PMTs}
R4998 PMTs have been delivered by Hamamatsu in  assemblies
(H6533) that include the PMT, the voltage divider chain
and a 1 mm $\mu-$metal shielding.
To increase the count rate stability of PMTs, instead of a conventional
resistive divider type, an active divider or a booster on the last dynodes
had to be used. 
After some tests, the performances of PMTs equipped with a booster or an
active divider were found roughly equivalent. The active divider 
option was chosen for its easier use.

About 120 H6533 assemblies were delivered by Hamamatsu in  two years.
 In the following, only the tests to study the rate capability
and the behaviour inside a  magnetic field will be shown. 
In addition, many tests were done to qualify the  PMT's assemblies 
for installation in the TOF detectors, as a sizeable fraction of them had
problems related to breakdown of the active divider under stress or showed 
a very noisy behaviour with big output spikes. 

\begin{figure*}[hbtp]
\begin{center}
\includegraphics[width=14cm]{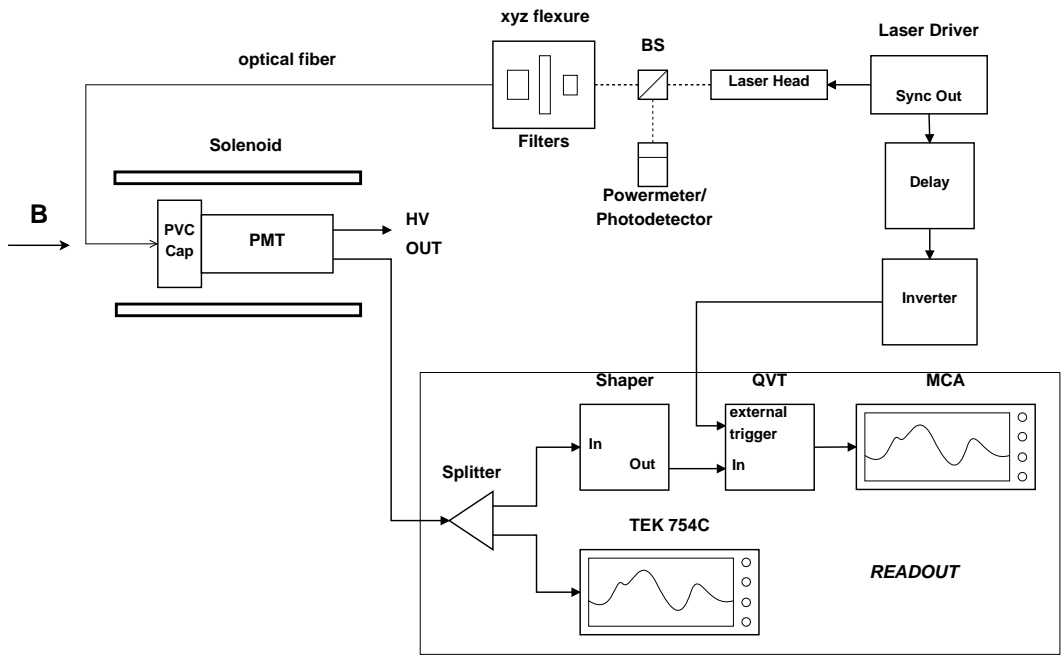}
\caption{ Scheme of the test setup for PMTs measurements (not in scale).
In some measurements
 the readout section (MCA) was replaced by a VME acquisition system,
based on a CAEN V2718 VME-PCI interface.}
\label{fig:setup}
\end{center}
\end{figure*}

To test single PMTs inside magnetic field or PMTs mounted on 
a scintillator bar a setup similar to the one on figure \ref{fig:setup} 
was used.
A fast light pulse~\footnote{
a home-made system
based on a Nichia NDHV310APC violet laser diode and an AvtechPulse fast
pulser (type AVO-9A-C laser diode driver, with $\sim200$ ps
risetime and AVX-S1 output module) was used.
This system gave laser pulses at $\sim 409 \ nm$
with a FWHM between $\sim 120$ ps and $\sim 3$ ns (as measured with a 6 GHz
6604B Tek scope)
and a max repetition rate $\sim 1$ MHz} 
was sent directly to the
PMT's photocathode via a 3 m long multimode 3M TECS FT-110-LMT optical
fiber (with a measured dispersion of $ \leq 15$ ps/m, see \cite{bonesini}).
At the end of the fiber a small Plexiglas prism, inserted in
a black plastic cover
in front of the PMT window, allowed illumination at the center of the
photocathode.
The laser spot was focused into the optical fiber (aligned by a micrometric
 x-y-z flexure system~\footnote{Thorlabs MBT613/M with 4 mm excursion and
a resolution of $\sim 0.5 \ \mu$m}) by a 10x Newport microscope objective, after  removable absorptive
neutral density filters, to give light signals of different intensities.
A broadband beamsplitter (BS) divided the laser beam to give $ 50 \% $
of light on the fiber injection system and $ 50 \%$ on a monitoring detector.
A fast Thorlabs DET210 photodiode (risetime $\sim 1$ ns) was used in most
measurements, to monitor the laser stability.
For gain measurements the PMT signal was acquired in average mode by
a Tektronix
TDS 754C digital scope (500 MHz bandwidth, 2 Gs/s sampling rate) 
triggered by the laser output 
syncronization signal (sync. out),   
 that had a maximum jitter of 15 ps with respect
to the delivered optical
 pulse. In part of the  measurements
 the signal was sent after a passive $50 \%$
T divider to a Canberra 2005 preamplifier,
followed by an EG-G Ortec 570 shaper (shaping time $\sim 1 \ \mu s$,
gain $\sim 200$)  followed by a Silena 8950 multichannel analyzer (MCA), 
using as external trigger the sync out signal of the laser.

For timing measurements, the same MCA chain
was used with a Silena 7422 charge-amplitude-time converter (QVT),  
see figure \ref{fig:setup} for details.
The STOP  signal ($t_{STOP}$) was given by the PMT anode signal after
a leading edge
PLS 707 discriminator, while the START signal ($t_{START}$) was given by the
sync out of the pulser after a suitable delay and an ORTEC  pulse inverter.
In timing measurements what is actually measured is the time difference
$\Delta t= t_{START}-t_{STOP}$, that accounts for delay in cables and
electronics and jitter in the transit time in the tested PMTs.
A lack of variation
in this quantity or no deterioration in the FWHM of its distribution,
after increasing the magnetic field intensity, demonstrates the effectiveness
of the adopted shielding.
The used TDC range (up to $0.1 \ \mu s$) with the MCA resolution (2K)
allowed a resolution of 50 ps/count.
\subsubsection{PMTs behavior in magnetic field}
Systematic studies have been done, using
a dedicated resistive solenoid of 23 cm inner diameter,
40 cm length~\footnote{
built by TBM,
Uboldo (VA), Italy}. 
The big open bore allows tests of single H6553 assemblies
both with field lines orthogonal or parallel to the
PMT axis up to $\sim 700$ Gauss.
The magnetic field was measured via a gaussmeter~\footnote{
Hirst GM04 model, with axial Hall probe},
with an accuracy better than $1 \%$. 
Tests were done usually with a signal corresponding to 
a MIP.
The laser optical power was periodically monitored with an
OPHIR NOVA laser power meter.
The number of photoelectrons ($N_{pe}$) was estimated via absolute
gain measurement.
This number was cross-checked with the power meter
measurements.
The PMTs were inserted in the central region of the test solenoid, where
the field had a uniformity better than  $ 3 \%$, using a support
 to incline them  between $0^0$ and $90^0$ with respect to the  field lines
in the magnet ({\bf B$_{\parallel}$ or B$_{\perp}$}). Environment light was
accurately masked to reduce noise.
 
\begin{figure*}[hbt]
\begin{center}
\includegraphics[width=5.5cm]{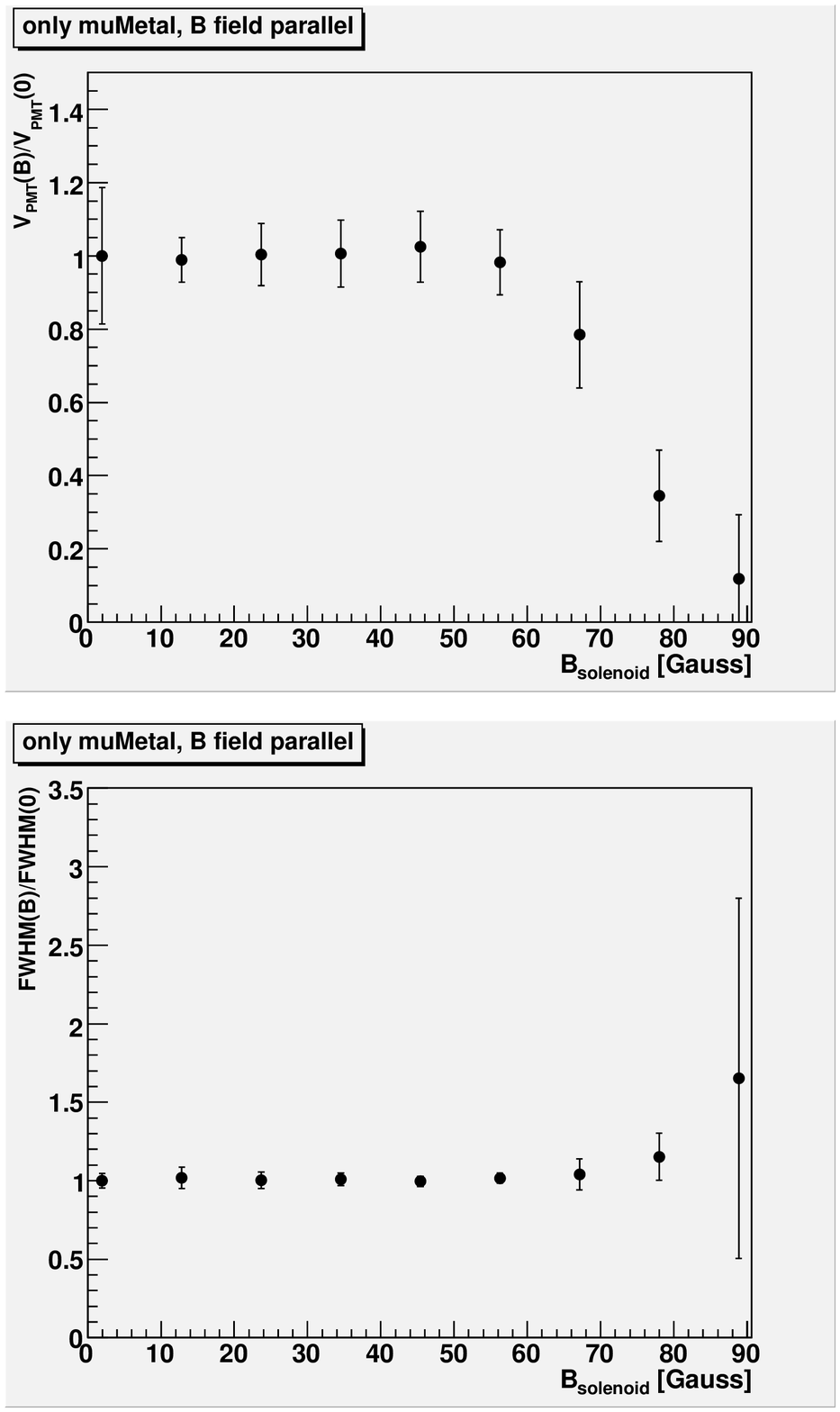}
\includegraphics[width=6.5cm]{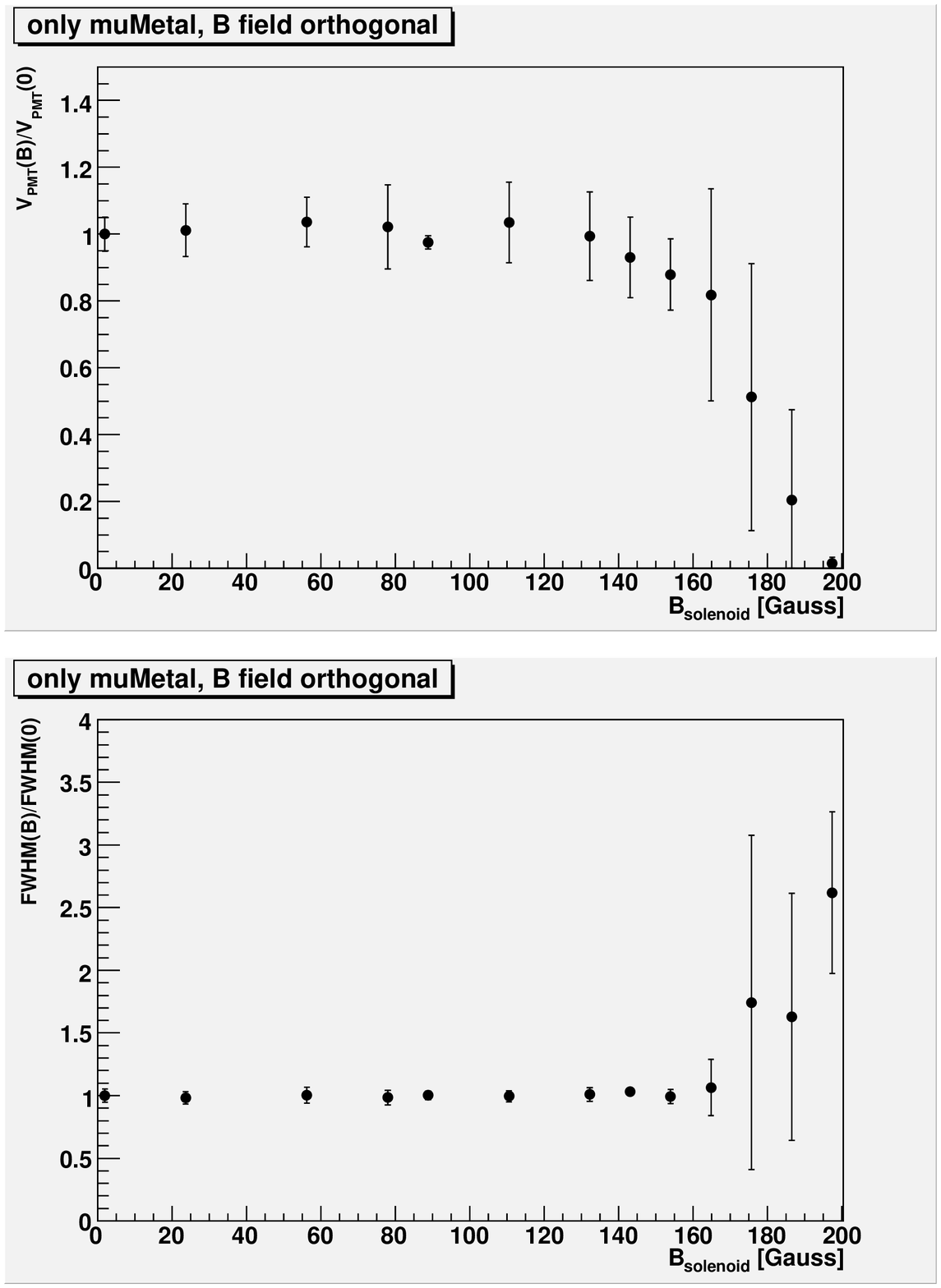}
\caption{ Signal ratio at field B and B=0 G and FWHM ratio at field B and
B=0 G for the timing difference, measured as $\Delta t=t_{START}-t_{STOP}$ with
only the mu-metal shielding of 1 mm for the PMTs. 
Left panel:  longitudinal field,
right panel: orthogonal field. The plots show the average and rms for a
sample of ten R4998 PMTs.}
\label{fig:test3}
\end{center}
\end{figure*}

Results for signal reduction and timing versus the
magnetic field intensity {\bf B} for the average and rms of a sample
of ten PMTs are shown  in figure \ref{fig:test3}.

The uncertainties in these studies came mainly from 
non-uniformity of the magnetic field,
stability of the laser pulses,
error in positioning of PMTs inside the magnetic field, 
conservatively  estimated to less than $10 \%$ and 
statistical errors. 

The  studies described above show that H6533 assemblies 
(with a 1 mm $\mu-$metal shielding) perform satisfactorily  inside
residual longitudinal magnetic fields up to $\sim 60$ Gauss and orthogonal
magnetic fields up to $\sim 150 $ Gauss. This is the case for TOF0
or TOF1 inside the external shielding cage.

\subsubsection{Rate capability}
\begin{figure}[hbt]
\begin{center}
\includegraphics[width=0.61\textwidth]{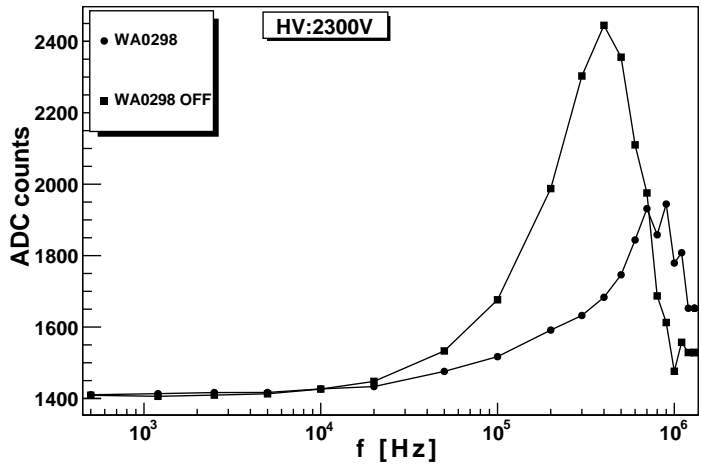}
\caption{ Effect of the booster for one PMT (WA0298) at a 
{\bf B=0 } G (signal in a.u. versus the rate f in Hz). The bottom line is with
the boster on}
\label{fig11}
\end{center}
\end{figure}

\begin{figure}[hbt]
\begin{center}
\vskip -2cm
\includegraphics[width=0.49\textwidth]{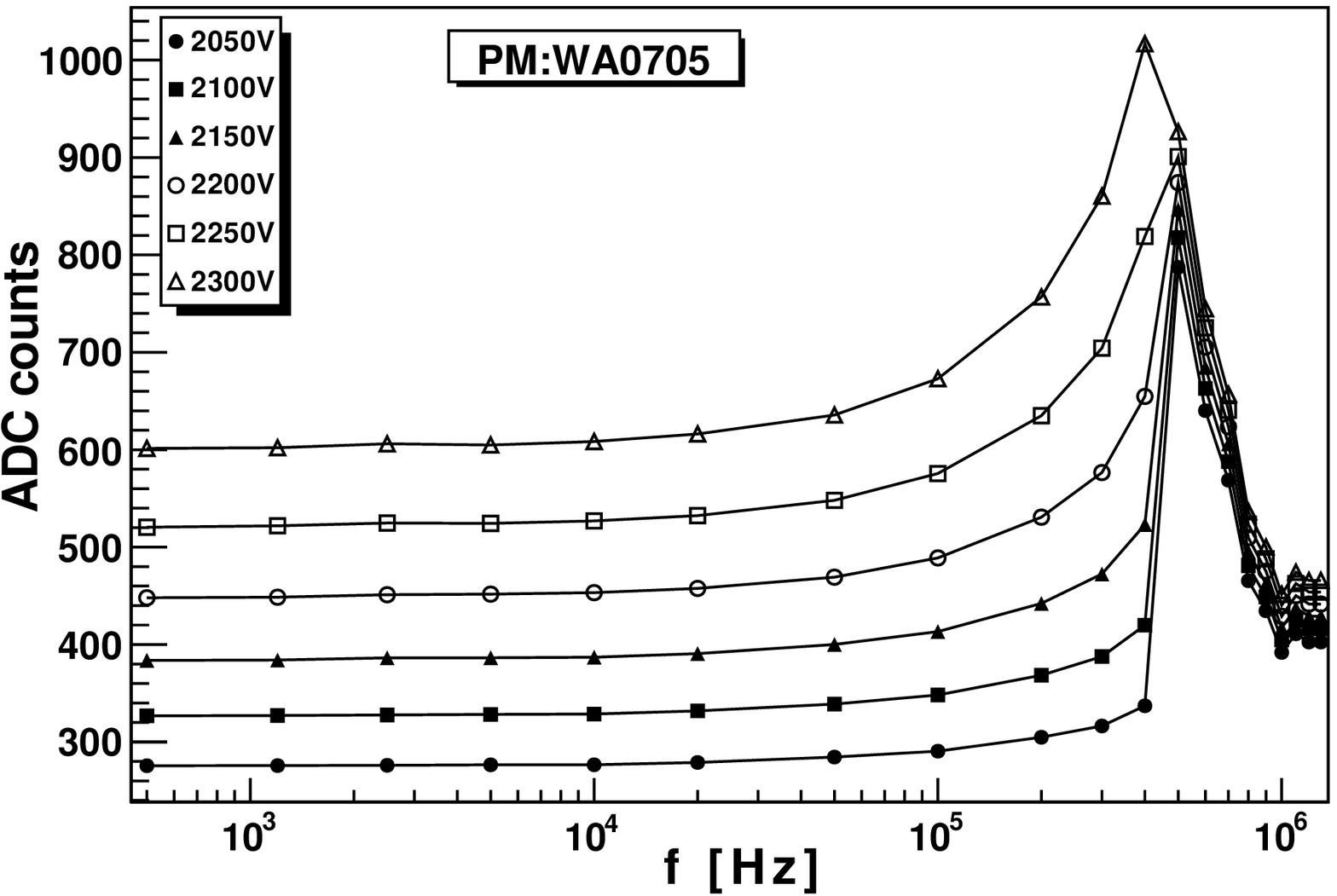} 
\includegraphics[width=0.49\textwidth]{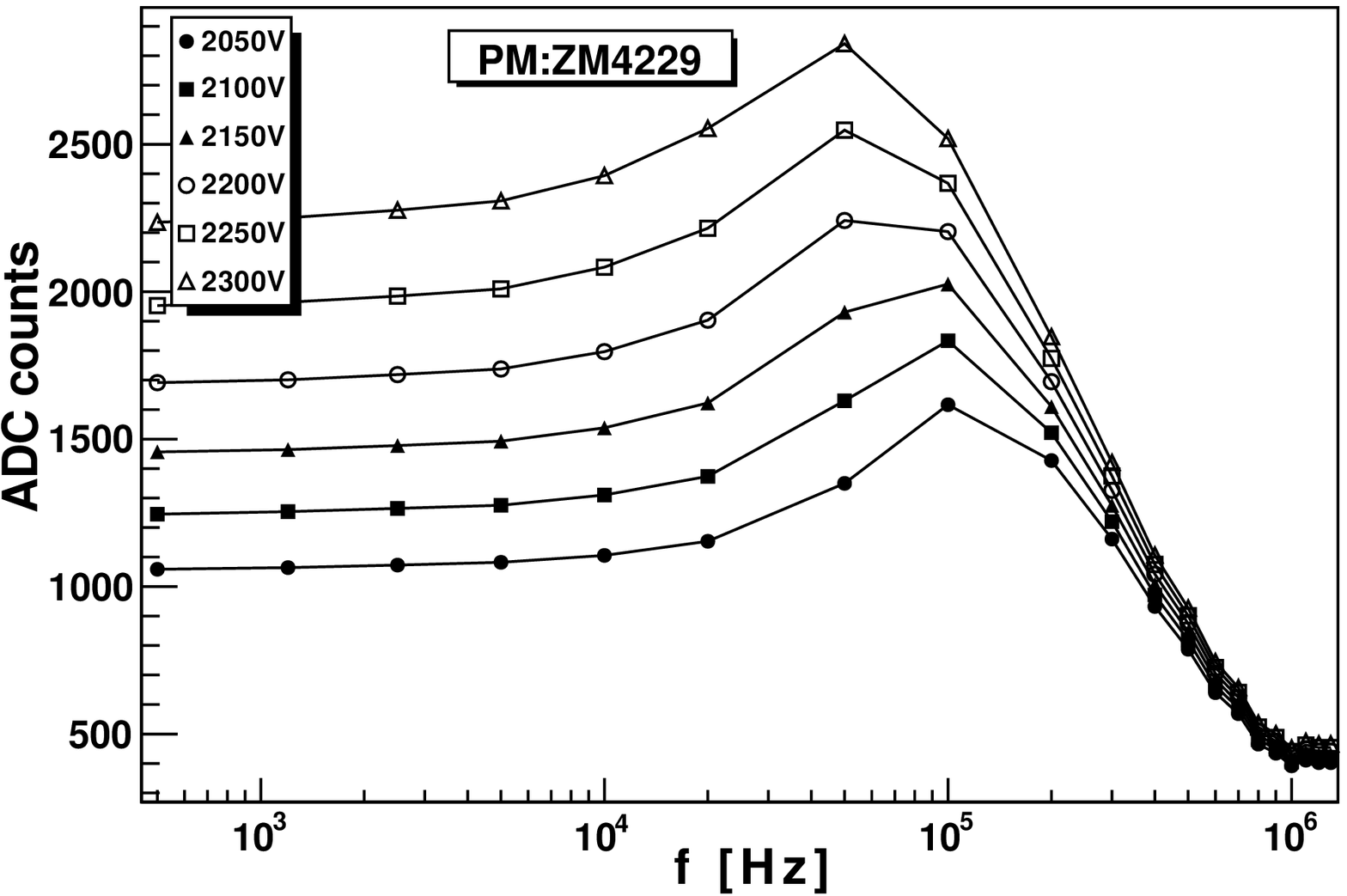} 
\includegraphics[width=0.49\textwidth]{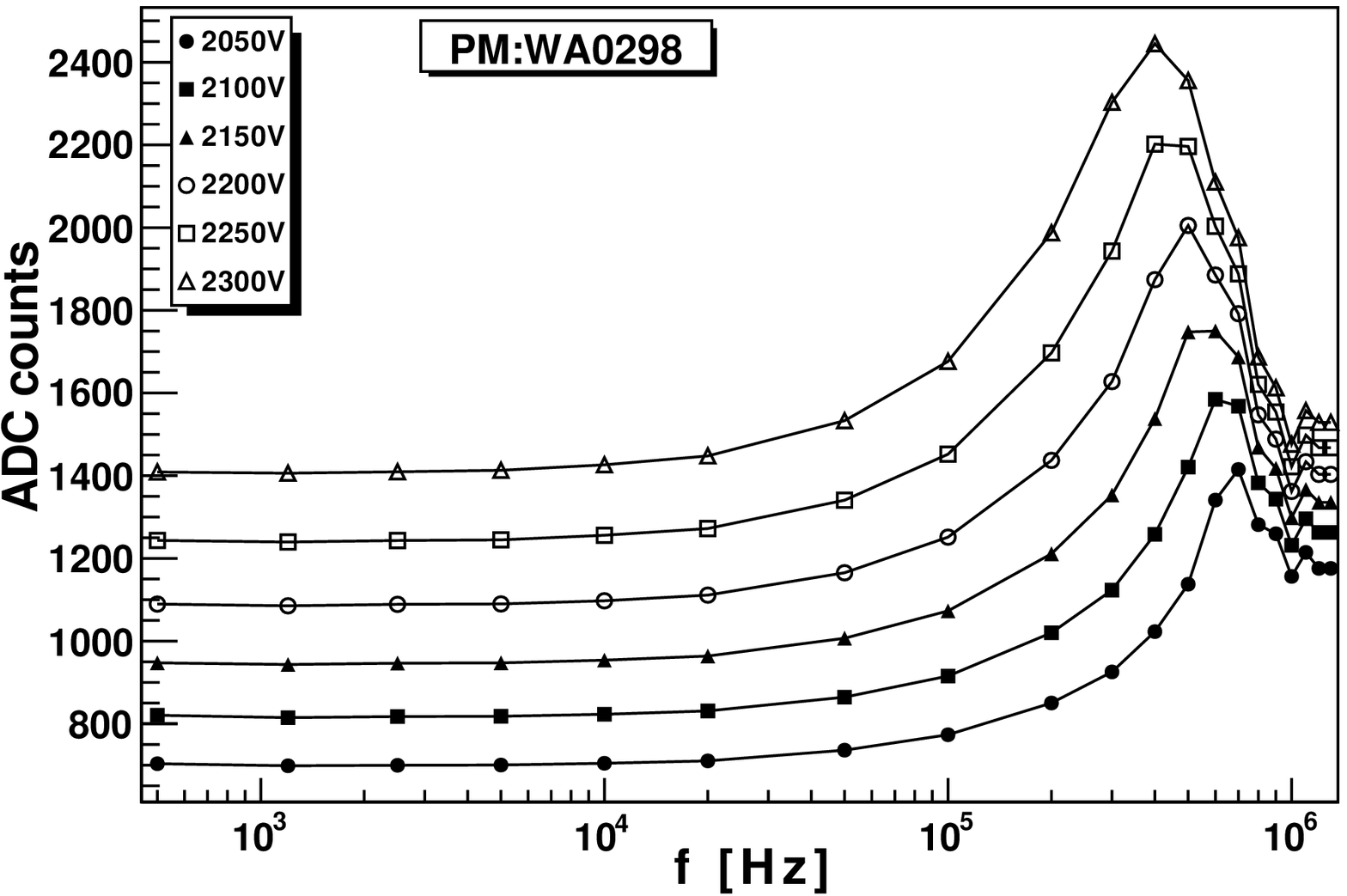} 
\includegraphics[width=0.49\textwidth]{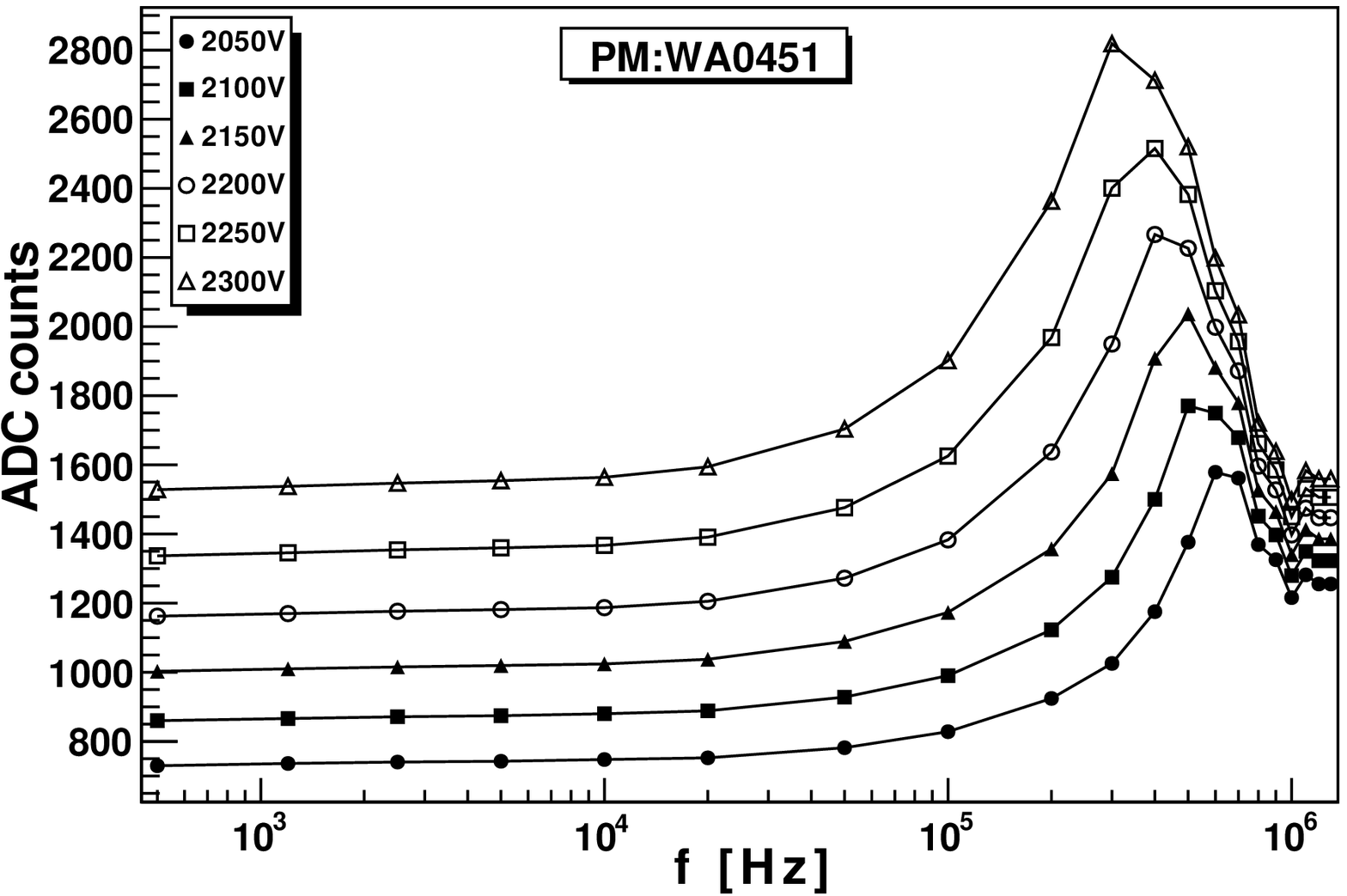} 
\includegraphics[width=0.49\textwidth]{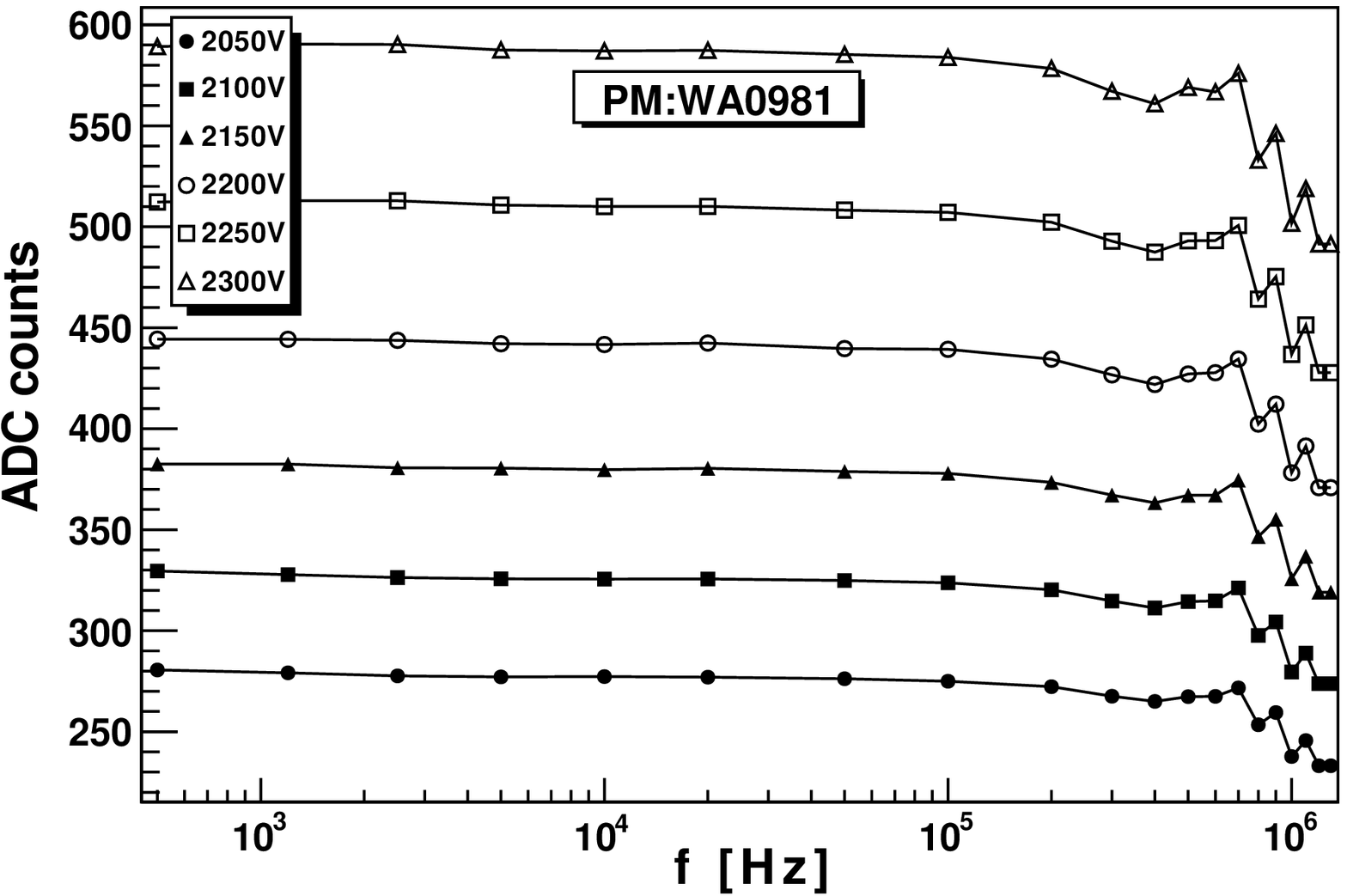} 
\includegraphics[width=0.49\textwidth]{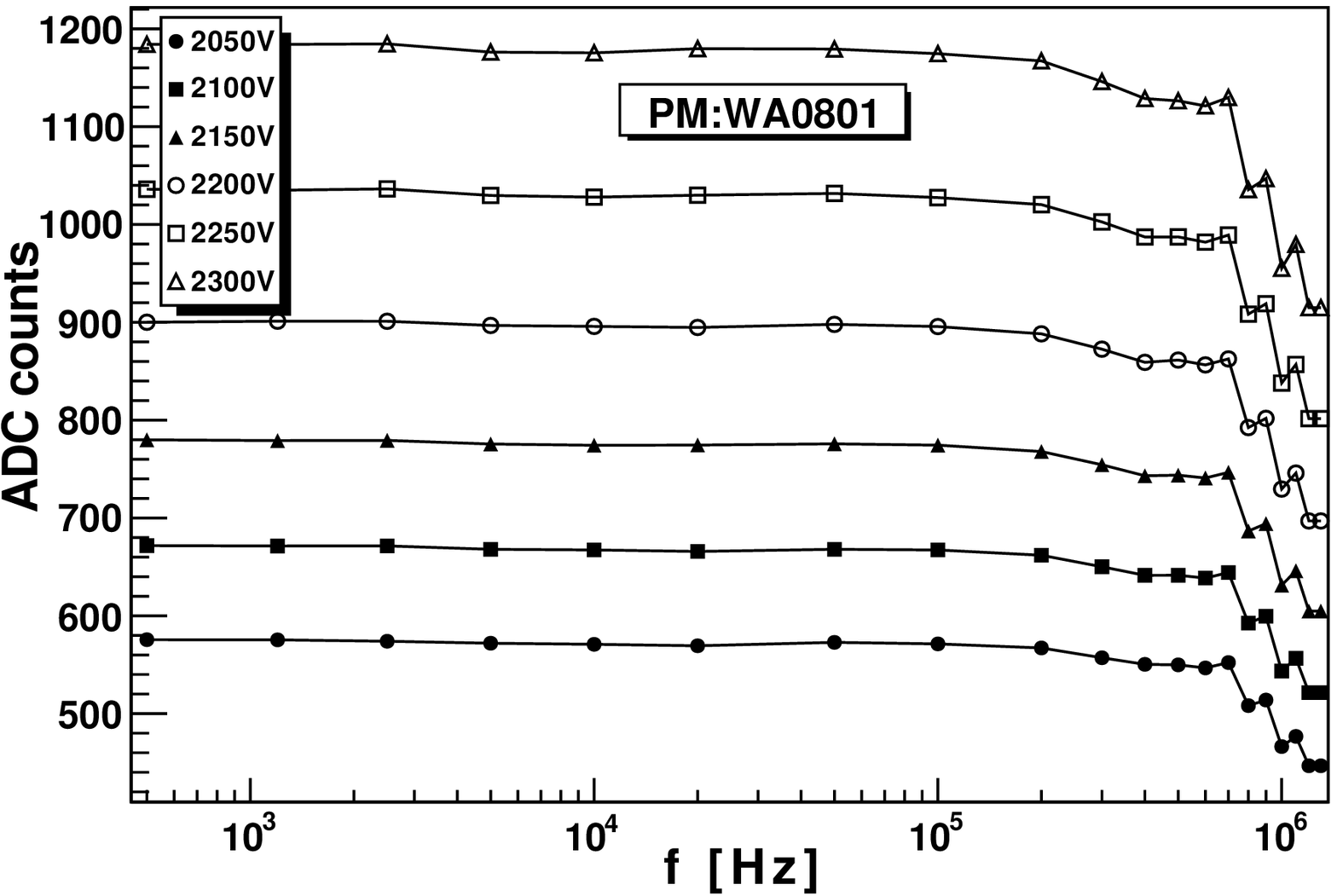} 
\caption{ Rate capability of typical R4998 PMTs, as a function of
rate R at field {\bf B=0 } G (signal in a.u. versus the rate f in Hz).
Top panels: with passive divider, middle panels: booster divider, 
bottom panels: with active divider. In each 
panel the H.V. is decreased in 50 V steps from -2300 V from 
top to bottom and a typical PMT with lower/higher gain is shown in the
left/right plot.}
\label{fig8}
\end{center}
\end{figure}

\begin{figure}[hbt]
\begin{center}
\includegraphics[width=0.49\textwidth]{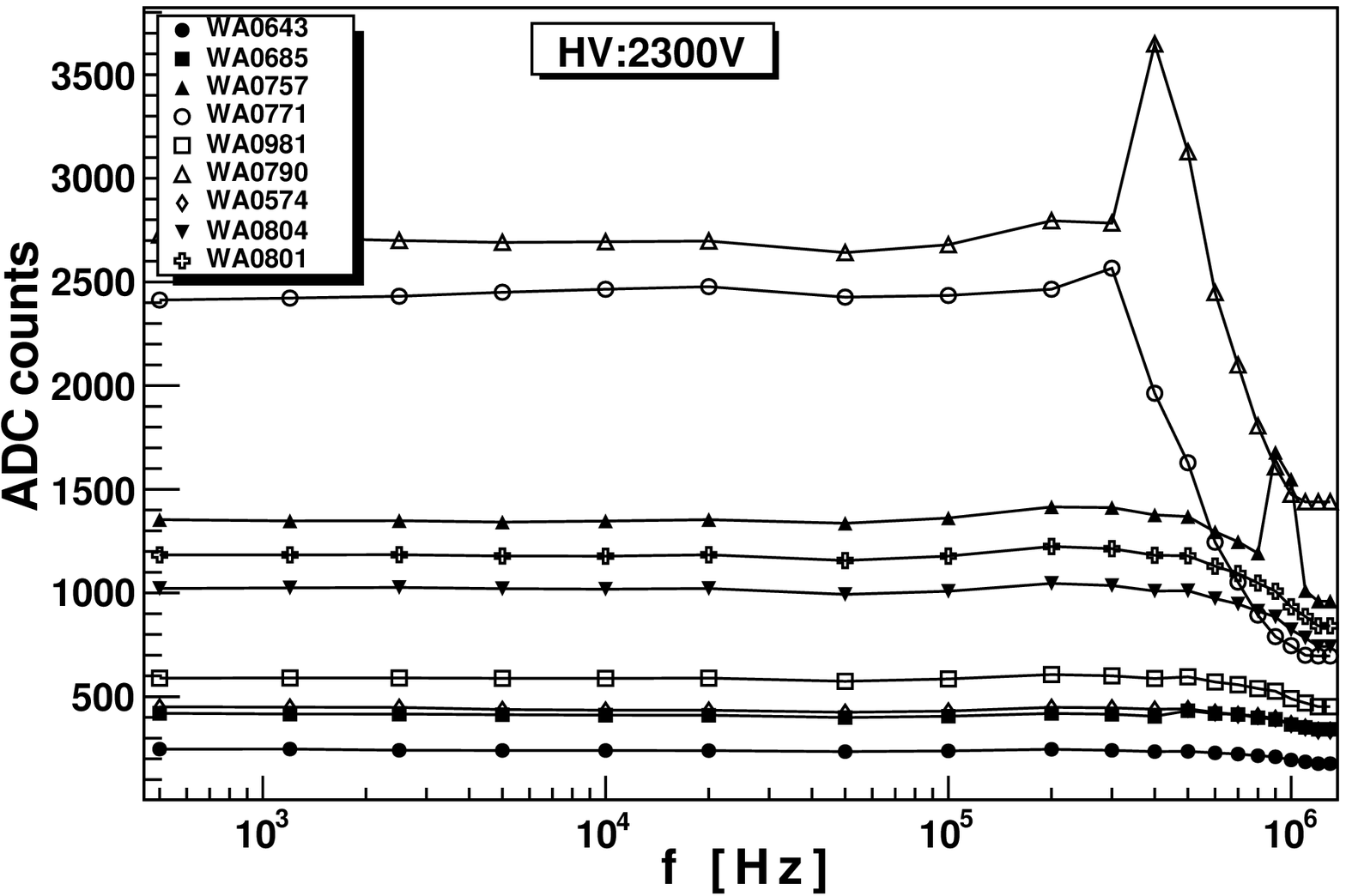} 
\includegraphics[width=0.49\textwidth]{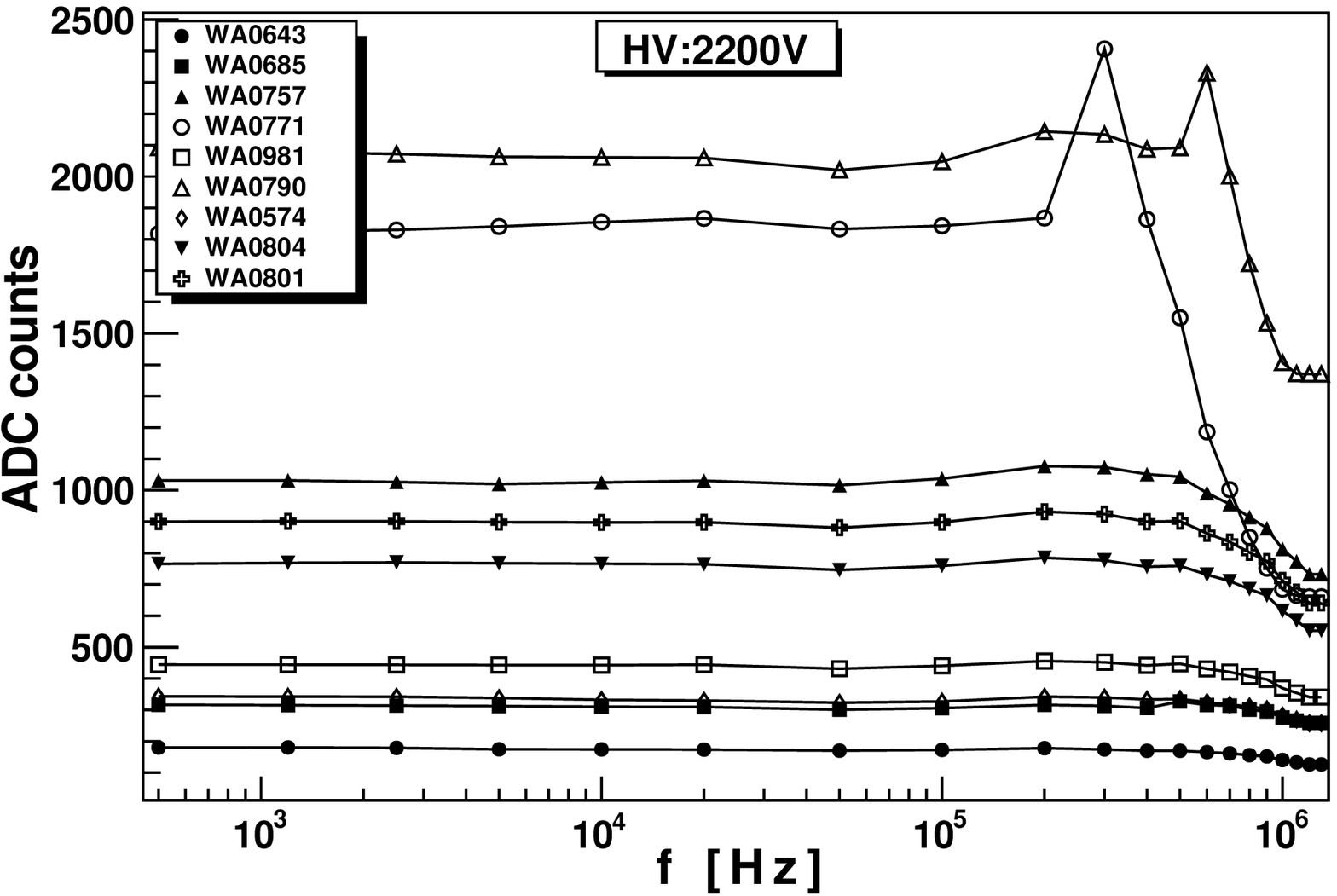} 
\includegraphics[width=0.49\textwidth]{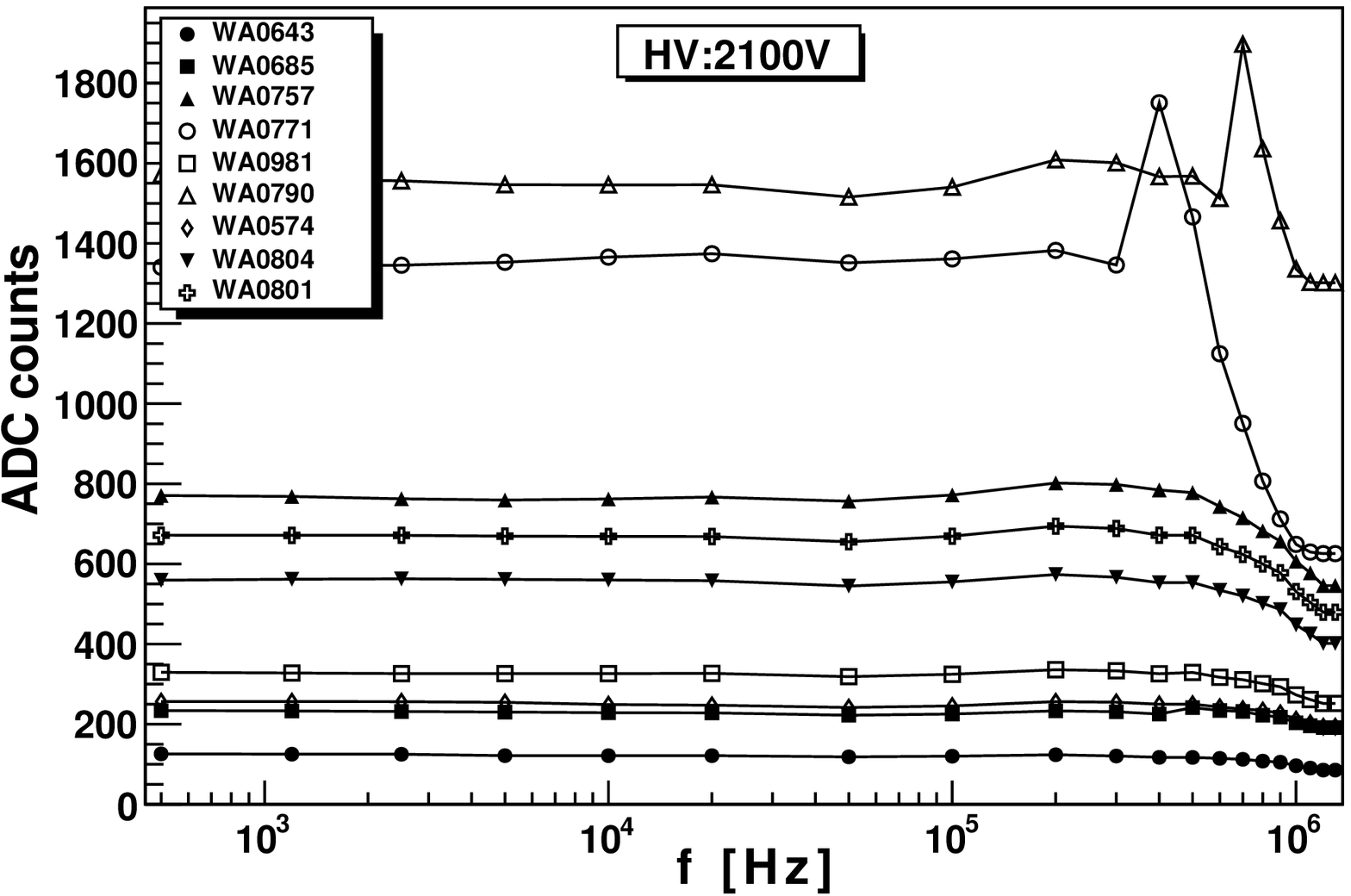} 
\includegraphics[width=0.49\textwidth]{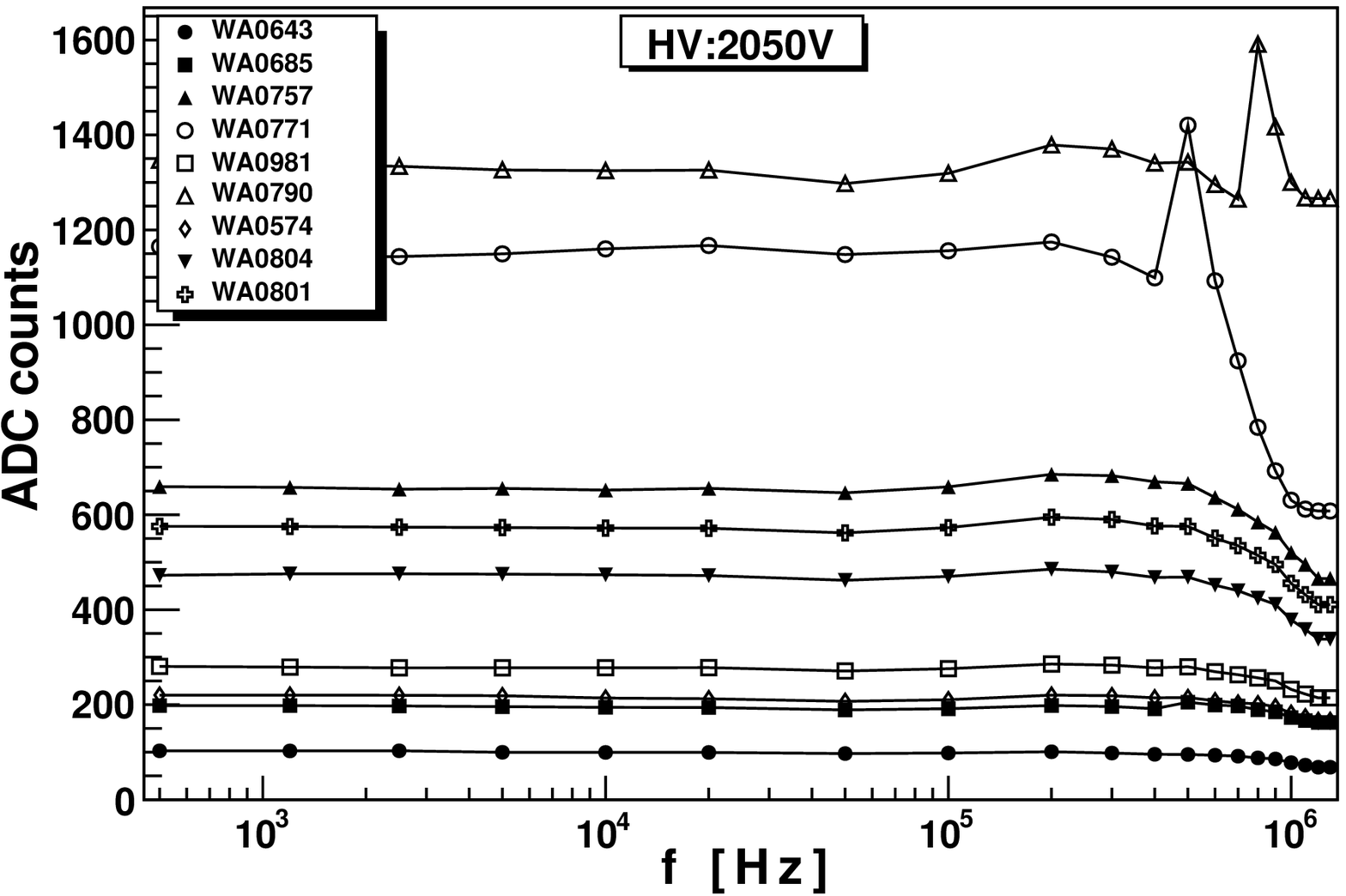} 
\caption{ Rate capability for a sample of nine R4998 PMTs with
active divider, as a function of
rate R at field {\bf B=0 } G (signal in a.u. versus the rate f in Hz).
The upper curves correspond to PMTS with very high gain (and noise) 
not used for detector readout.}
\label{fig9}
\end{center}
\end{figure}

Complete scintillator counters equipped with PMTs at the ends
were used in these measurements. The laser light was injected 
in the scintillator bar through the standard laser 
injection system described in section 3.
The Avtech pulser was triggered externally, while the PMTs signals were digitized
by a CAEN V792 QADC and acquired by a CAEN V2718 PCI-VME  interface. 
The effect of a booster on the last dynodes for a typical PMT is shown
in figure \ref{fig11}.
Figure \ref{fig8} shows the PMT amplitude response (in a.u.) as a function
of the laser shot repetition rate R (simulating an increasing particle rate), 
 both with a conventional
resistive divider, a booster on the last dynodes  or an active divider
for typical PMTs. The HV is decreased 
in 50 V steps from the value of -2300 V at the top.
The laser light signal is 
roughly equivalent to one MIP for the curves at -2300 V.
These results can be easily understood, recalling that the rate capability
is limited by the maximum allowable anode current $I_a$,
that depends also on the PMT gain. 
The left panels are relative to PMTs with lower gains, as respect to the 
ones in the right panels (roughly a factor 2-3).
The rate capability for a sample of nine PMTs is shown instead in figure 
\ref{fig9} for some typical HV settings.

\begin{figure}[hbt]
\begin{center}
\includegraphics[width=6cm]{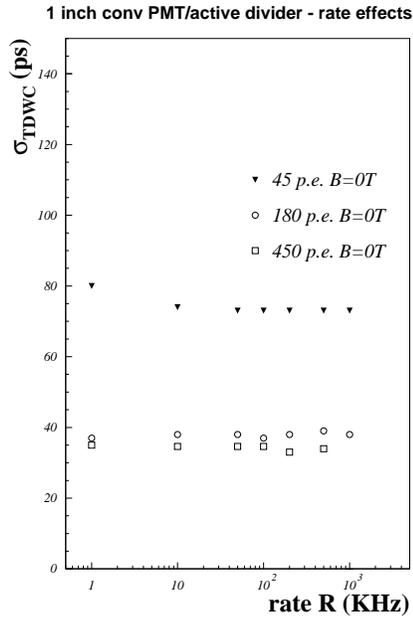}
\caption{ Timing resolution 
in  ps  as a function of the 
rate R (at B=0 T) for a one R4998 PMT with active divider. For a MIP,
signals correspond typically to the open symbols in the lower part
of the figure.}
\label{fig10}
\end{center}
\end{figure}

Timing characteristics of R4998 PMTs show no deterioration going from 1KHz up to 1 MHz, 
for sensible number of photoelectrons as seen in figure \ref{fig10} at {\bf B=}0 Gauss
for a typical assembly with active divider.

As a conclusion, from the performed laboratory tests, it appears that R4998 PMTs
may sustain rates up to 500-600 KHz without major problems with
an active divider or a booster, while this limit goes down 
to $\sim 100-200$ KHz with a resistive divider, depending on
the PMT gain. 

\subsection{Tests on single counters at the BTF facility}
The Laboratori Nazionali di Frascati (LNF) DA$\Phi$NE Beam Test Facility (BTF) is a beam transfer line designed to
deliver electrons or positrons mainly for detector calibration
purposes \cite{btf}.
Tests at the BTF were  done to choose the scintillator
to be used, cross-check the design of the lightguides, assess 
the counter intrinsic time resolution and measure the number 
of produced photoelectrons ($N_{p.e.}$).

The BTF pulse time structure allows to test resolution effects 
and eventually pile-up effects, but
not the behaviour with high rates.
Most of the data were taken in single particle mode (one electron per
pulse) exploiting  a high resolution calorimeter as diagnostic device.

The prototype TOF counters under study were mounted on a test stand
at a distance of about 10 cm one
from the other, with  
two finger counters (F1 and F2) of $5 \times 20$ mm$^2$  transverse area
to define the impinging beam. 
 
As scintillator material Bicron 
BC404, BC420 and BC408 and Amcrys-H UPS95F were used (see Table \ref{tab1}), while lightguides were
made of Bicron BC800, REPSOL Glass UVT PMMA or PERSPEX UVA PMMA.
In some measurements the 
left/right PMT signal
was used as START/STOP for a time-to-amplitude (TAC) unit 
connected to a MCA~\footnote{
ORTEC Trump-8K MCA with an ORTEC 566 TAC and a PLS711 leading-edge 
discriminator}, to get an immediate feedback on time resolutions. 

In the bulk of data taking each PMT signal was split by an active or
passive splitter to both a QADC line~\footnote{CAEN V792 model, 
32 channels, 12 bit, 0.1 pC/ch} 
 and , after a discriminator, to a TDC line.
By an appropriate fan-in, fan-out the baseline CAEN V1290 TDC was used together with a conventional TDC~\footnote{
CAEN V775 model, 12 bits, nominally $\sim 35$ ps/ch} for reference.
To convert TDC counts to picoseconds, the  V1290 TDC has
been  later calibrated offline with known delay cables.
A value 22 ps/count has been obtained, to be compared to
a nominal value of 25 ps/count.   
The adoption of leading edge discriminators (such as CAEN N417 or PLS 711)
introduced a time-walk effect.

Before data taking, the gains of the two PMTs of a given prototype counter 
were roughly equalized with cosmics looking at the signals
on a digital scope.

Event selection required a coincidence from the two
finger counters ($F1 \cdot F2$) and a pulse height compatible
with single impinging electron.  By fitting a gaussian
to the distribution of  $(t_L - t_R)/2$, with $t_{L/R}$ being the arrival
time at the $L/R$ PMTs of a single counter  as measured by the TDC chain, 
 it was possible to obtain
the counter intrinsic resolution $\sigma_t$. 

\begin{figure}[hbt]
\begin{center}
\includegraphics[width=8cm,height=4cm]{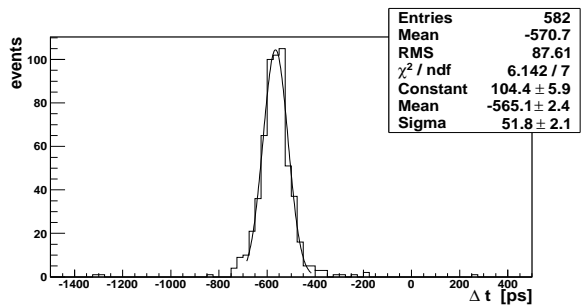}
\caption{$(t_L-t_R)/2$ distribution from a specimen BC404 bar and beam impact point at x=20 cm
(counter centre).}
\label{figBC404}
\end{center}
\end{figure}

Effects of the time-walk cancel out when the beam impinge on the
 center of a counter
and both PMTs give similar pulse heights.
Figure \ref{figBC404} shows, as an example, 
the  distribution of  $(t_L - t_R)/2$  for runs 
with beam hitting the center of one BC404 bar.
Table \ref{tab2} shows  the results obtained  for several prototype
counters with the beam hitting the counter centre.
Intrinsic time resolutions are all in the range  45-60 ps,
with $\sim20 \%$ better resolutions for BC420 or BC404 counters.
Similar results were obtained also by using fine-mesh one-inch Hamamatsu R5505 
PMTs in place of the conventional 
one-inch R4998 PMTs \footnote{R5505 PMTs have a TTS of $\sim 350$ ps and a nominal
gain of $\sim 5 \times 10^5$ at +2000 V}.

\begin{table}[hbt]
\begin{center}
{\footnotesize
\begin{tabular}{|c|c|}
\hline
counter type  & $\sigma_t$ (ps) \\ \hline
UPS95F 4cm bar Winston Cone      & $56 \pm 2$ \\
UPS95F 4cm bar REPSOL UVT lightguide & $50 \pm 8$ \\
BC404  6cm bar REPSOL UVT lightguide & $ 46 \pm 5$ \\
BC420  6cm bar REPSOL UVT lightguide & $ 45 \pm 1$ \\
BC408  6cm bar PERSPEX UVA lightguide& $ 60 \pm 2$ \\
\hline \hline
\end{tabular}
\caption{Intrinsic resolution of counters made of scintillation bars
  of 4 or 6 cm width and with lightguides made of different materials
  and/or of different shape (Winston cone or fishtail).}
\label{tab2}}
\end{center}
\end{table}

\begin{figure}[hbt]
\begin{center}
\includegraphics[width=0.40\linewidth]{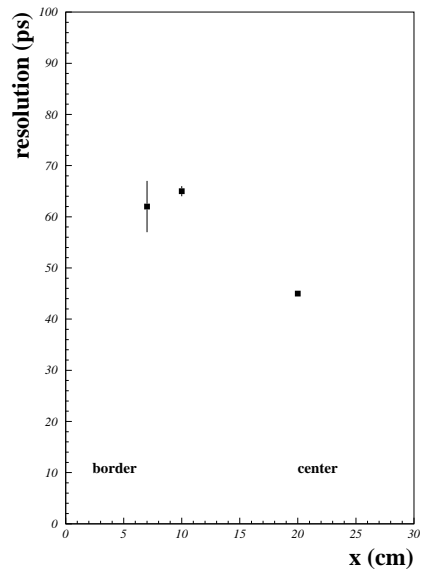}
\caption{Resolution in ps for a 40 cm long, 6 cm  wide BC420 scintillation counter, as a function
of the impact point x in cm (x=20 cm corresponds to the center of the counter).}
\label{fig-scan}
\end{center}
\end{figure}

When scanning along a counter,  effects of PMT non-equality and from
time-walk~\footnote{this last effect may be corrected for with a pulse height
measurement, using a time-walk correction} show up as demonstrated in figure \ref{fig-scan}.

If the pulse height distribution is fully described
by the photoelectron statistics,  it is possible to estimate 
the number of photoelectrons per single impinging electron ($N_{p.e.}$) from:

\begin{equation}
N^{raw}_{p.e.} \simeq (\frac{<R>}{\sigma_{R}})^2 
\label{ref:npe}
\end{equation}

where the average pulse-height $<R>$ and the resolution $\sigma_{R}$ are
obtained from a gaussian fit. 

This estimation neglects electronic noise and gain fluctuations and is
affected by the quality of the scintillator-PMT coupling.

As explained in reference \cite{astro},  the estimation uncertainty of formula
(\ref{ref:npe}) depends on the amplification factors of the first and second
dynodes. For R4998 PMTs these factors  are estimated as $\sim 10$ and
$\sim 3-5$\cite{confalonieri}, giving 
a correction factor of about $\sim + 10\%$ for $N^{raw}_{p.e.}$. 
From the available data $N_{pe}$ is estimated to be in the range of 200-300 p.e.
for the BC420 counters under test, depending on the impact beam position.

The number of photoelectrons can be also estimated on simple grounds 
with the formula:

\begin{equation}
N_{pe}= \frac{dE/dx (MeV/cm)}{h\nu(eV)} \times \epsilon_{scint} \times t(cm)
\times \epsilon_{opt} \times Q.E.
\label{eq:npe}
\end{equation}
where $\epsilon_{scint}$ is the conversion efficiency of deposited energy
into scintillation photons (usually $\sim .01$), $t$ is the scintillator
thickness in cm, $Q.E.$ is the PMTs photocathode quantum efficiency and 
$\epsilon_{opt}$ is the optical collection efficiency,
to be estimated by simulation.
The light collection in the TOF counter has been simulated with the
program GUIDEIT \cite{guideit}, using light sources uniformly dispersed along
the median crossing plane of the counter.
From the simulation the collection efficiency has been estimated 
to be $\sim 3.8 \%$ and from formula \ref{eq:npe} it can be estimated
that $N_{pe} \sim 230 $ p.e. in agreement with  the previous estimation based on photo-electron statistics.\\

\section{Detector commissioning at RAL}

To equalize the amplitude response of the TOF0 and TOF1 scintillation
counters~\footnote{This is useful in order also to have similar time-walk
corrections for the timing response of the two (L/R) PMTs of the same
counter} the different PMT's gain and
the optical coupling \footnote{ both between the scintillator bar
and the lightguides and the lightguide collars and the PMTs}
 in the L/R side of each bar must be accounted for.
Neglecting this last factor, a pre-equalization has been done by 
taking into account only to the PMT's gains.
By using a 
YAP:CE source 
from SCIONIX Ltd.~\footnote{ with a
nominal rate of $\sim 20$ counts/s, a calibration run of a few $10^3$ events
was done in about 10-20 minutes instead of the many days needed in a
cosmics testbench} pulse height spectra were recorded both on a digital
scope~\footnote{Tektronix DPO7054, with a 2.5 GHz bandwidth} and
with a VME CAEN V792 QADC, read by a CAEN V2718 VME-PCI interface.

\begin{figure}[hbt]
\begin{center}
\includegraphics[width=0.4\textwidth]{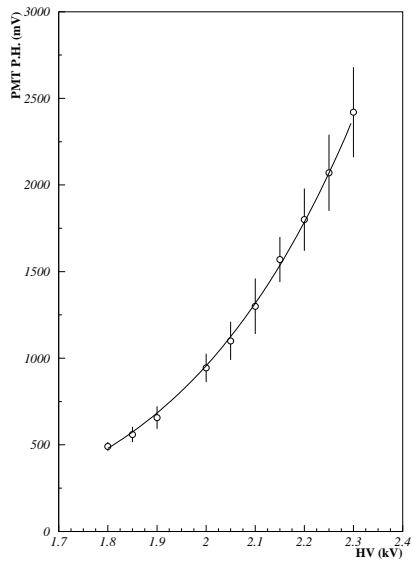}
\caption{ Dependence of average P.H. from H.V. settings, for a
typical Hamamatsu R4998 PMT.}
\label{fig:fit}
\end{center}
\end{figure}

\begin{figure}[hbt]
\begin{center}
\includegraphics[width=0.9\textwidth]{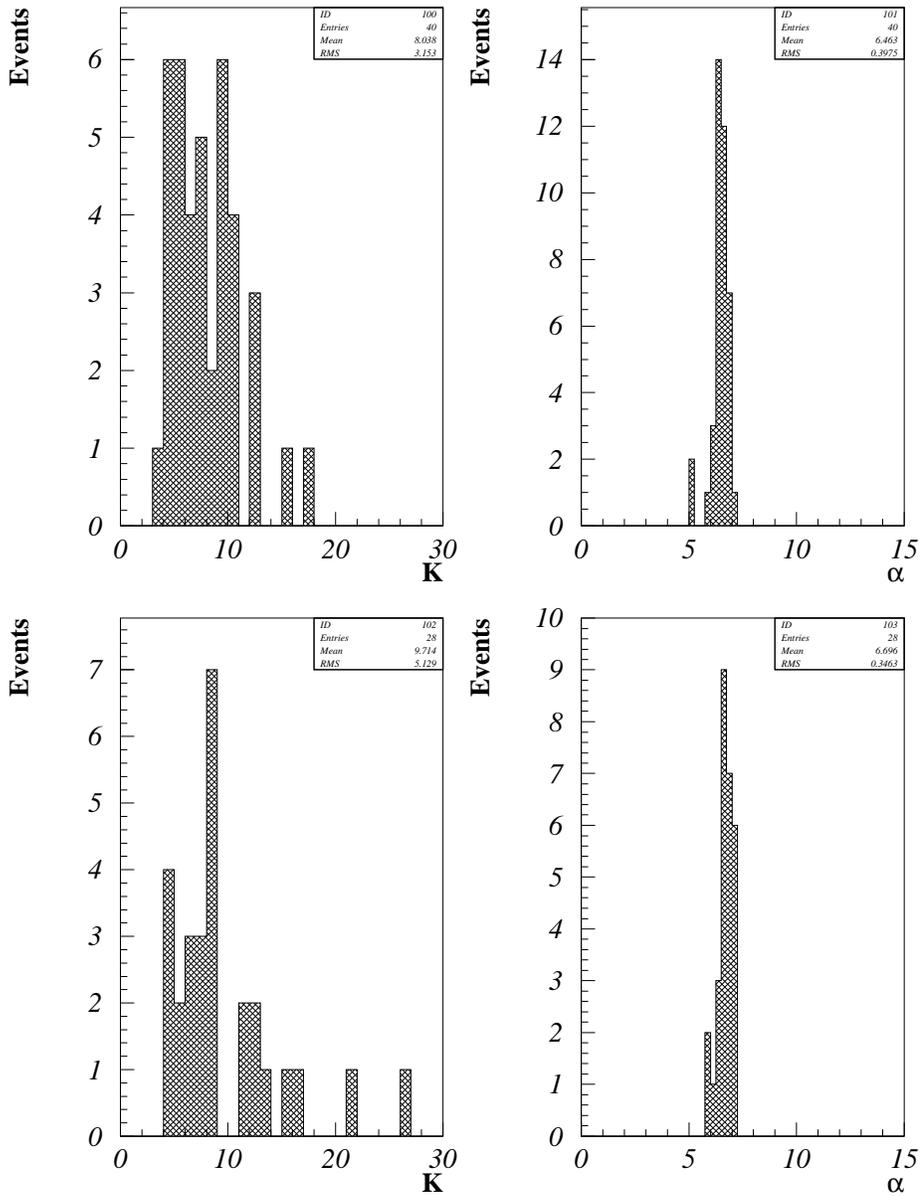}
\caption{ Distributions of the fitted parameters K and $\alpha$  for the PMTs
used in the TOF0 (upper panels) and TOF1 (lower panels) detectors.}
\label{fig:fitx}
\end{center}
\end{figure}

Data were recorded at a nominal H.V. value, set on a CAEN N470 module 
from about -1800 V to -2300 V, in 50V steps.
The amplitude (in mV) has been plotted as a function of the H.V.
 (in kV) and fitted with a functional form $K \times V^{\alpha}$,
with $K, \alpha$ free parameters for each PMT.  Figure \ref{fig:fit}
shows the fit for a typical PMTs.
Figure \ref{fig:fitx} shows instead the distribution of the $K$ and
$\alpha$ parameters for the sets of PMTs used in TOF0 (upper panels)
and TOF1 (lower panels) detectors. 
The $K$ and $\alpha$ values of each PMT and the functional form
$K \times V^{\alpha}$ were then  used in the detector
equalization for amplitudes.
While the ``normalization'' $K$ parameter varies up to a factor of ten
with a mean value 8.04 (9.71) and  $r.m.s.$ of 3.15 (5.13) for TOF0
(TOF1), the $\alpha$ parameter giving the ``slope'' of the correction
has a mean value 6.46 (6.69) with $r.m.s.$  of 0.40 (0.35) for TOF0
(TOF1).

In the scintillation counters pre-calibration procedure, 
the PMT working voltages
 have been set, trying to select the left and right PMTs  of each counter 
with the most similar gains
and the PMTs for the horizontal and vertical planes of the same detector
with similar gains.

\subsection{First performance in beam}

Put outside the closed DSA area~\footnote{Decay Solenoid Area - closed area 
nearby the extraction point
of the pion secondary beam from ISIS that contains a 5 m long, 5 T decay 
solenoid for muon collection
and the first PID detectors, including TOF0}, on a special
trolley, TOF0 was tested 
in July 2008  
to assess PMTs reliability in real
working condition during the summer 2008 ISIS run and then moved to its
final position inside DSA in September 2008.
TOF1 has been installed instead at RAL in December 2008 on the 
temporary trolley after TOF0 
and since then tested with a few dedicated runs.
Due to problems in the cooling of the decay solenoid, that persistently affected
its performance, only some low intensity runs with positrons or pions 
were done to test preliminary detector performance. 

For a particle crossing a scintillation counter $i$ (i=1, ...10),
equipped with two photomultipliers  $j$ (j=1,2) of a plane
$l$ (l=1,2) of a TOF detector, at a time $t_0$ and at a distance 
$x$ from its center,
the signal arrival time at the PMT photocathode $t_{i,j,l}$ is given by:
\begin{eqnarray*}
t_{i,j,l}=t_0 + \frac{L/2 \pm x}{v_{eff}} + \delta_{i,j,l}
 \ \ \ \ \ \ \ \
 j=1,2;  l=1,2 
\label{eq:deltat}
\end{eqnarray*}
where {\em L} is the scintillator length, $v_{eff}$ the effective light velocity
in the scintillator slab  and
 $\delta_{i,j,l}$ include
all delays (cables, PMT transit time, etc.).
After correction for the delays $\delta_{i,j,l}$, the quantity
\begin{eqnarray*}
 t_{+,i,l}=\frac{ t_{i,1,l} +  t_{i,2,l}}{2}=t_0+\frac{L}{2
\cdot v_{eff}}-t_s 
\label{media}
\end{eqnarray*}
 is independent of the impact point $x$ along the counter $i$ and
allows  measurement of the time-of-flight (TOF) in a detector plane, 
while the impact position $x$ can be deduced from
\begin{eqnarray*}
t_{-,i,l}=\frac{t_{i,1,l} - t_{i,2,l}}{2}=\frac{x}{v_{eff}}. \nonumber
\label{diff1}
\end{eqnarray*}

The calculation of the delays $\delta_{i,j,l}$ is a quite delicate task 
and may be done with impinging beam particles. 

For the trigger TOF0 detector, defining as a ``pixel'' the 
area given by the crossing of 
two orthogonal slabs $i,k$ (in the horizontal and the vertical plane of a TOF
detector), the calibration procedure first determines the peak position
of timing signals with respect to the trigger~\footnote{ for an 
incoming particle
the trigger signal is given by the first of the twofold coincidences 
from slab $i$ and slab $k$. The time of the coincidence signal is the
time of the latest signal arriving to the logic unit.}, 
for particles hitting a pixel.
From these, 
alignment time 
calibration constants may be deduced.
For the second TOF station (TOF1), in the calibration procedure one has to
account also for the additional delay due to the time-of-flight between the 
two stations, using particles of known velocity (such as positrons).   

\begin{figure}[hbt]
\begin{center}
\includegraphics[width=8cm]{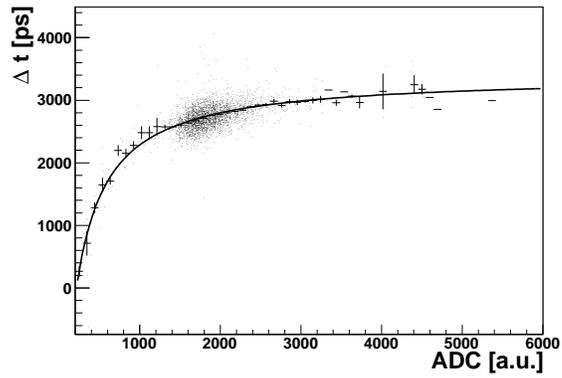}
\caption{Time walk and fitted function for a typical PMT.}
\label{tw}
\end{center}
\end{figure}

The adoption of leading edge discriminators (such as Lecroy 4415) introduces
a dependence of the discriminating threshold crossing time on the collected charge
(time-walk).
To correct for time-walk, the dependence of the difference 
between the time measured by the TDC and a reference time on the maximum 
of the signal of the PMT, as measured by the FADC, the following function is fitted to the data:
\begin{eqnarray*}
f(ADC) = P_{1} + \frac{P_{2}}{(ADC+ P_{0})} + 
\frac{P_{3}}{(ADC + P_{0})^{2}}
\end{eqnarray*}
with parameters $P_0, P_1, P_2, P_3$ determined for each PMT,
as shown in figure \ref{tw}.

The reference time is given by a PMT in the other plane of the 
station. 

\begin{figure}[hbt]
\begin{center}
\includegraphics[width=7cm]{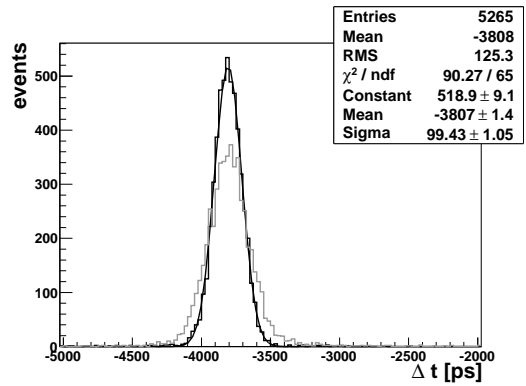}
\caption{The time difference between the slab 4 in plane 0  and  slabs 5
in plane 1 before and after $time \ walk$ correction.}
\label{twresol}
\end{center}
\end{figure}

The precision in the calculation of the time-walk correction is limited by the 
very poor collected  statistics for $ADC$ under 1000 counts 
and above 3000  counts (see figure \ref{tw}).
The effect of the time-walk correction is illustrated in figure
\ref{twresol} for a typical counter.

The calculation of the time calibration constants was done exploiting 
300 Mev/c pion beam data and after the time-walk 
correction. The collected data were just enough for calibration of only 
9 central pixels in TOF0 and 2 central pixels in TOF1.

\begin{figure}[hbt]
\begin{center}
\includegraphics[width=9cm]{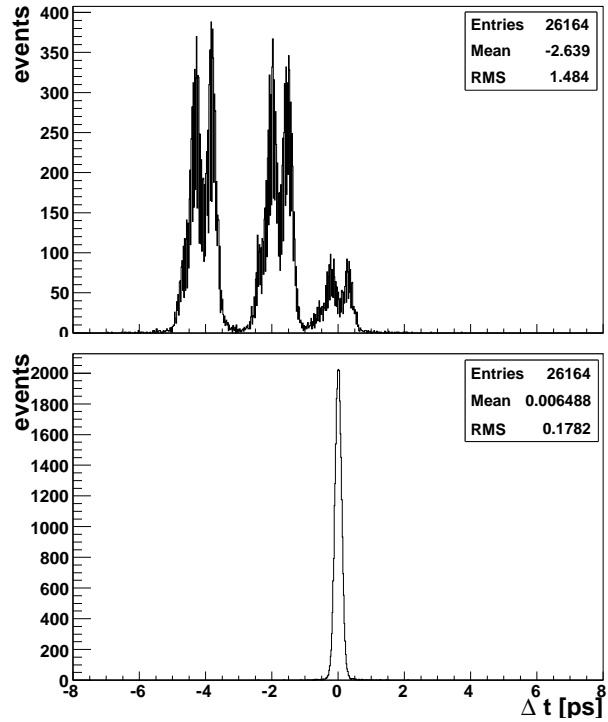}
\caption{Distribution of the time difference between the vertical and 
horizontal slabs for all the counters in TOF0 without (top) and with (bottom) 
the time corrections. Only events in the "pixels" where statistics allowed
calibration were considered.}
\label{calib}
\end{center}
\end{figure}

The effect of the absolute time calibration and  the time  walk 
correction   is illustrated in figure \ref{calib}.

\begin{figure}[hbt]
\begin{center}
\includegraphics[width=7cm]{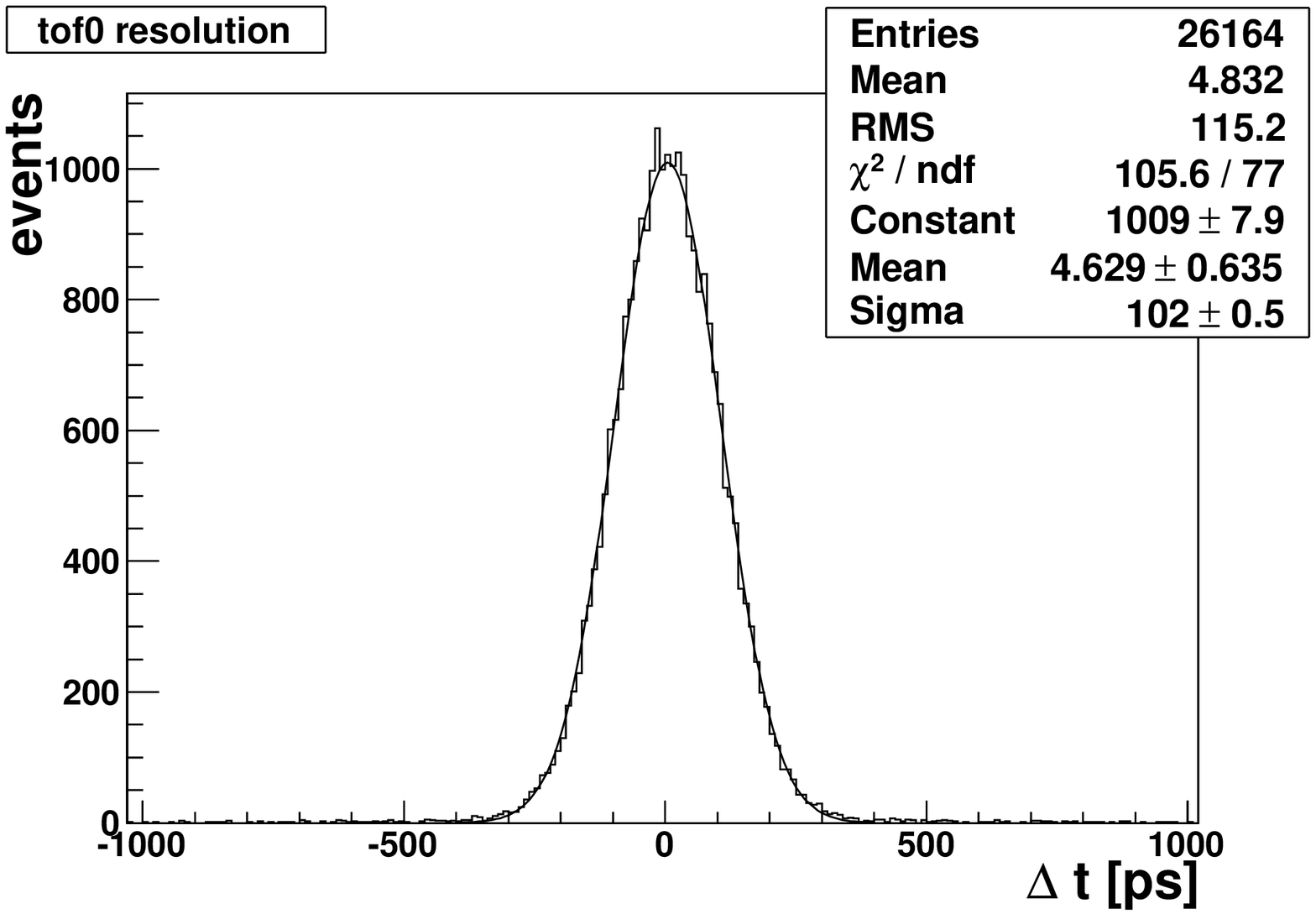}
\includegraphics[width=7cm]{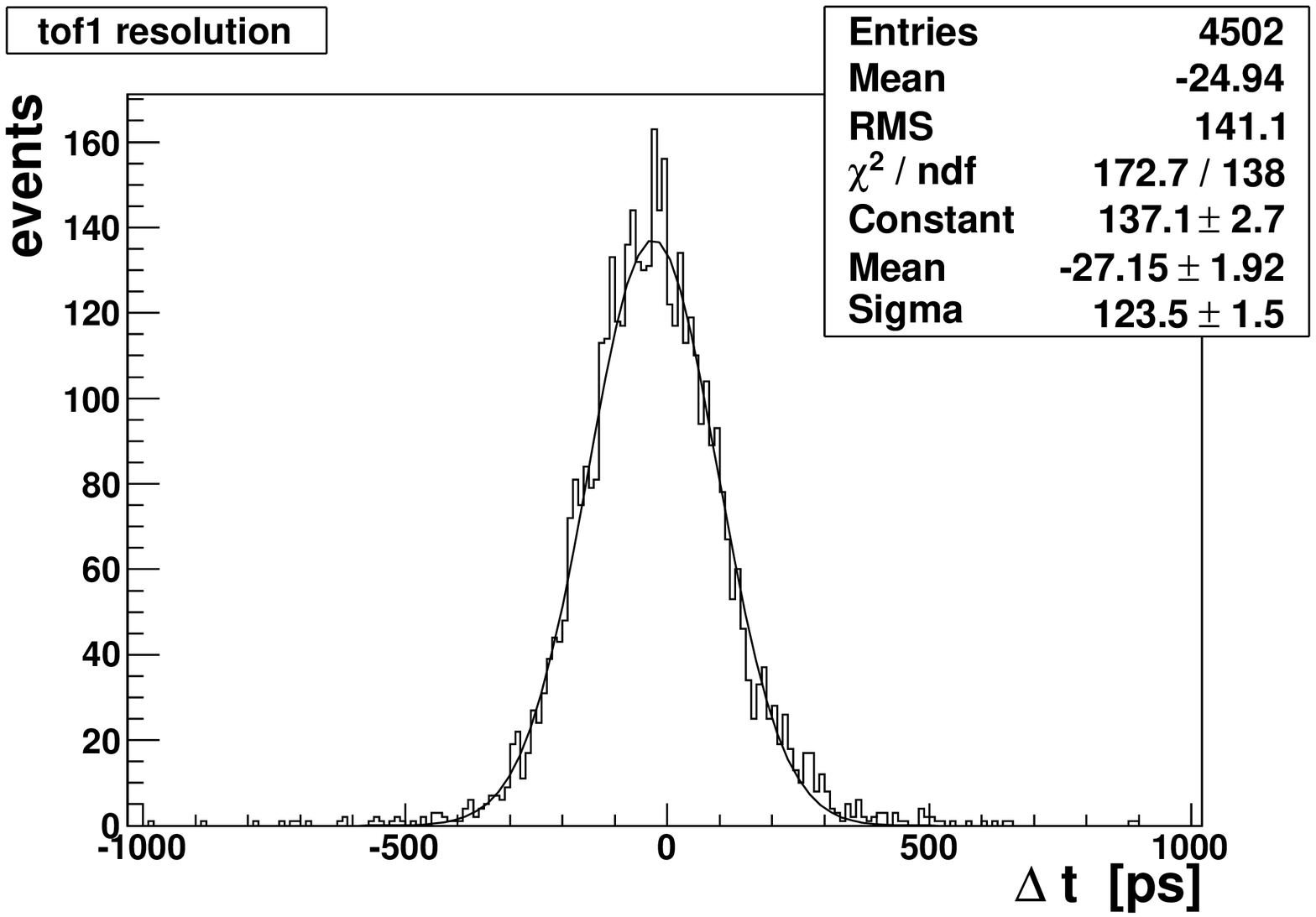}
\caption{Top (bottom) panel: time difference $\Delta t_{XY}$ between the vertical and
horizontal slabs in TOF0 (TOF1).}
\label{tofresol}
\end{center}
\end{figure}

The resolution after the calibration can be measured by using the time
difference $\Delta t_{XY}$  between the vertical and horizontal slabs in the same TOF
station (see figure \ref{tofresol}).
The obtained resolution on the difference is $\sigma^{0}_{XY}\sim 102 \ ps$ 
for TOF0 and
$\sigma^{1}_{XY} \sim 123 \ ps$ for TOF1\footnote{This translates into 
$\sim 50 (60) ps$
resolution for the full TOF0 (TOF1) detector with crossed horizontal and vertical slabs. }.

\begin{figure}[htb]
\begin{center}
\includegraphics[width=7cm]{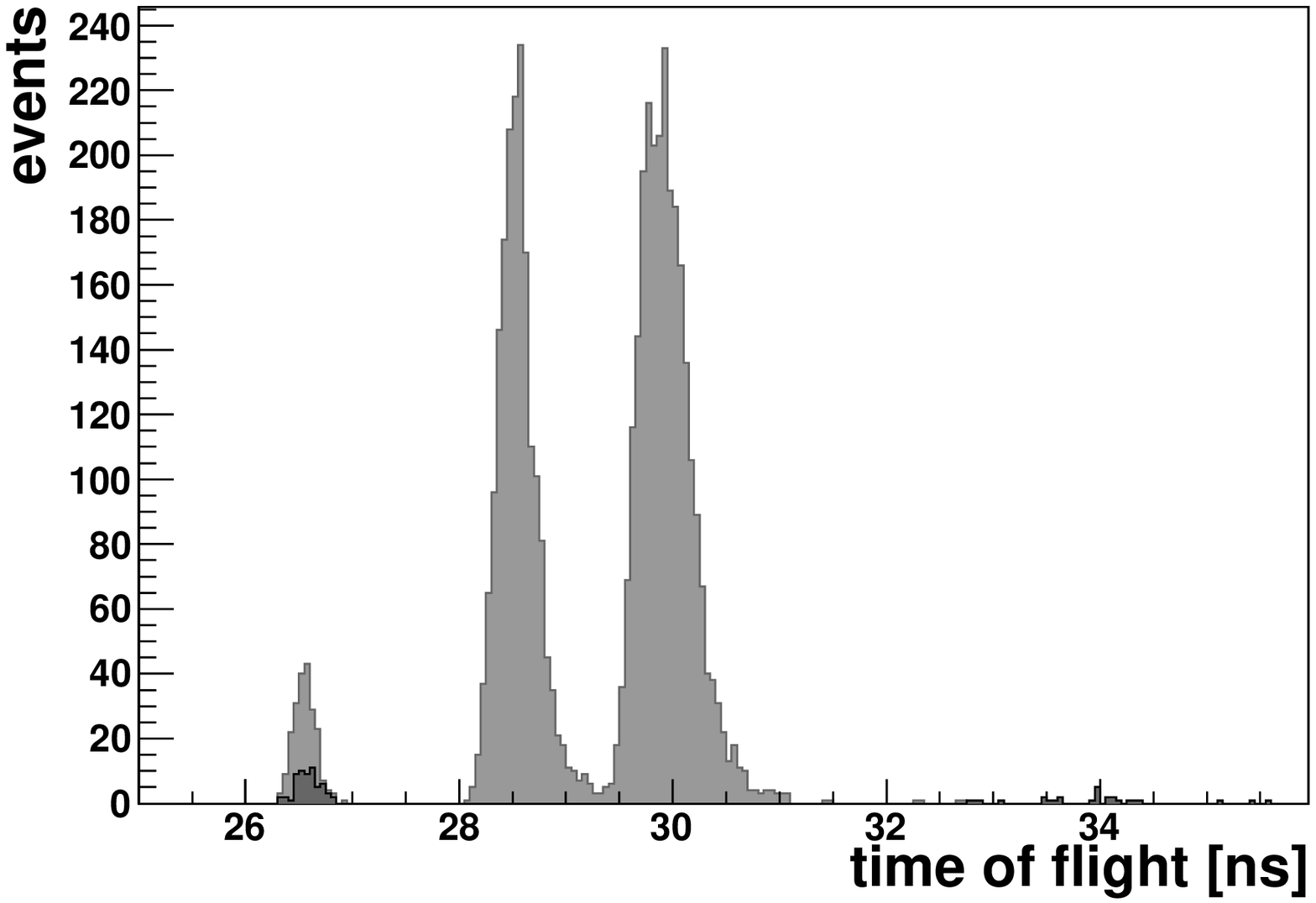}
\includegraphics[width=7cm]{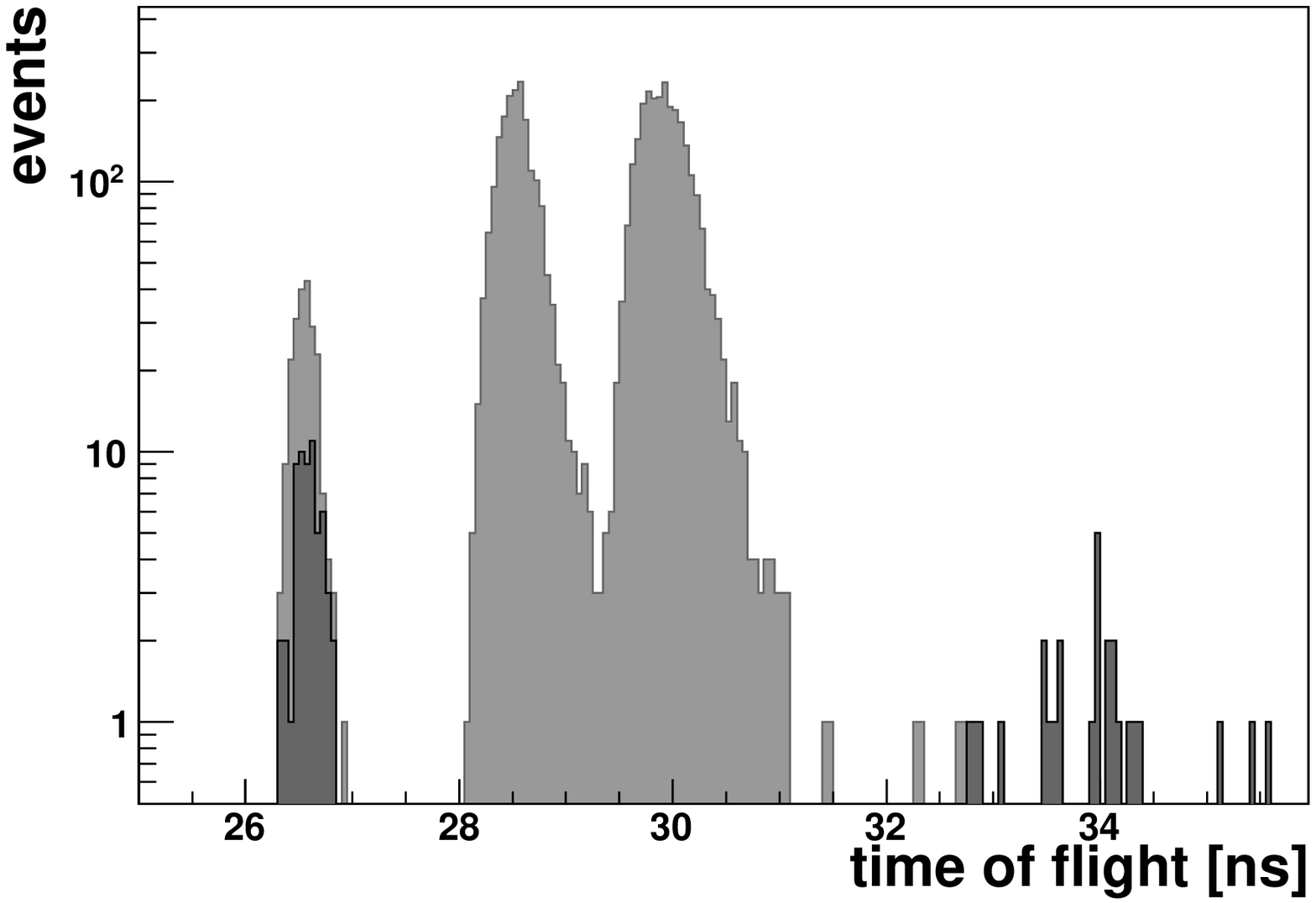}
\caption{ Time of flight between TOF0 and TOF1 for the  $positron$
(black) and $pion$ (grey) beams in normal (top) and logarithmic (bottom) scale.}
\label{tof}
\end{center}
\end{figure}

Figure \ref{tof} shows the distribution of the
time-of-flight between TOF0 and TOF1
for the 300 MeV/c pion beam 
and  a positron beam~\footnote{this beam is set by starting from 
the settings for pion beam at 300 MeV/c and reducing down
all the currents in the upstream magnets to a nominal
100 MeV/c momentum. At this momentum only positrons reach TOF stations}.
The first peak which is present in both distributions (pion and
positron beam) is considered as the $time \ of \ flight$ of the positrons
and is used to determine the absolute value of the time in TOF1.
A natural interpretation of the other two peaks is that they are due
to forward flying muons from pion decay and pions themselves.

\section*{Conclusions}
This paper reports the design and commissioning of the upstream section of
the MICE time-of-flight detector system and preliminary evaluation of its  
performance in the beam. After a calibration with impinging particles,
an intrinsic detector  resolution of $\sim50-60  $ ps is obtained.
A TOF measurement between two stations with a resolution 
of $\sim70-80 $ ps is thus within reach. 

\section*{Acknowledgements} 
We acknowledge the essential help of Mr. R. Mazza of INFN Milano
Bicocca for the skilful design of TOF0/TOF1 mechanics and Mr. 
S. Banfi, R. Gheigher from INFN Milano Bicocca and T. Locatelli,
C. Scagliotti and A. Freddi from INFN Pavia for their contributions
to the construction. 
We are grateful to all MICE collaborators for useful
discussions and encouragement in the couse of this work, in particular
to J. Cobb, G. Gregoire, W. Lau and L. Tortora. 
These measurements were carried out using the purpose-built MICE beam-line
at the ISIS facility at the STFC Rutherford Appleton Laboratory. It is a
pleasure to acknowledge the efforts of many people, from ISIS and the MICE
collaboration, in developing and operating the MICE beam-line, and the
ongoing support of the host facility.

\clearpage

\clearpage 



\clearpage
\begin{thebibliography}{99}
\bibitem{mice} A. Blondel et al., proposal, RAL, 2003.
\bibitem{kosharev} D.G. Kosharev, CERN/ISR-DI/74-62 (174); \\
 A. Blondel et al., CERN-2004-002; \\
M. Bonesini,A. Guglielmi Phys. Rep. 433 (2006) 65.
\bibitem{tosca} Tosca-2D or Tosca-3D programs from Vector Fields Inc.,
 http://www.vectorfields.com
\bibitem{comsol} COMSOL Multiphysics from COMSOL Inc., 
http://www.comsol.com 
\bibitem{cobb} J.Cobb, H.Witte, private communication
\bibitem{gregoire1} G. Gregoire, private communication
\bibitem{gregoire} G. Gregoire, W. Lau, private communication
\bibitem{bonesini2} M. Bonesini et al., Nucl. Instr. Meth. A567 (2006) 200; \\
M. Bonesini et al., Nucl. Instr. Meth. A572 (2007) 465.
\bibitem{guideit} D.A.Simon, Guideit v 1.1 Manual, 1993  
\bibitem{bonesini}M. Bonesini et al., IEEE Trans. Nucl. Science {\bf 50} (2003) 541 
\bibitem{harp-tof} M. Baldo-Ceolin et al., Nucl. Instr. Meth. A532 (2004) 548.
\bibitem{tintori} C. Tintori, HPTDC workshop, CERN, 2003 
\bibitem{btf} G. Mazzitelli, A. Ghigo, F. Sannibale, P. Valente, G. Vignola,
Nucl. Instr. Meth. A515 (2003) 524; \\
B. Buonomo,G. Mazzitelli and P. Valente, IEEE Trans. Nucl. Sc. 
NS-52(4) 2005 824.
\bibitem{astro} T. Yamaoka et al., 28th Int. cosmic Ray Conf., proceedings, 
p. 2871, 2003.
\bibitem{confalonieri} L. Confalonieri, Hamamatsu, private communication.
\end{thebibliography}
\end{document}